# Two-loop electroweak corrections to the $\rho$ parameter beyond the leading approximation

**Joannis Papavassiliou and Kostas Philippides**

Department of Physics, New York University,
4 Washington Place,
New York, N. Y. 10003, USA

and

**Ken Sasaki**

Department of Physics, Yokohama National University,
Yokohama 240, JAPAN

ABSTRACT

We show that in the framework of the pinch technique the universal part of the $\rho$ parameter can be meaningfully defined, beyond one loop. The universal part so obtained satisfies the crucial requirements of gauge-independence, finiteness, and process-independence, even when subleading contributions of the top quark are included. The mechanism which enforces the aforementioned properties is explained in detail, and several subtle field theoretical issues are discussed. Explicit calculations of the sub-leading two-loop corrections of order $O(G_\mu^2 m_t^2 M_Z^2)$ are carried out in the context of an $SU(2)$ model, with $M_W = M_Z$, and various intermediate and final results are reported.

# 1. Introduction

One of the most important quantities of the standard model is the $\rho$-parameter [1], defined as the ratio of the relative strength between neutral and charged current interactions, at low momentum transfer, namely

$$\rho = \frac{G_{NC}(0)}{G_{CC}(0)} = \frac{1}{1 - \Delta\rho} \ . \tag{1.1}$$

The $\rho$-parameter displays a strong dependence on $m_t$ and affects most electroweak parameters such as $\Delta r$, $M_W$, and $\sin^2\theta_{eff}(M_Z)$. The $\rho$ parameter defined above as the ratio of two amplitudes is a gauge independent and finite quantity. In addition, it is manifestly process dependent, and its value depends on the quantum numbers of the external particles chosen. To fully determine the value of $\rho$ for a given neutral and charged process, one must compute the complete set of Feynman diagrams ( self-energy, vertex and box graphs) to a given order in perturbation theory. However, traditionally one focuses instead on the quantity $\Delta$, defined as

$$\Delta = \frac{A_W(0)}{M_W^2} - \frac{A_Z(0) + (2s/c)A_{\gamma Z}(0)}{M_Z^2} \ , \tag{1.2}$$

where $A_W$, $A_Z$ and $A_{\gamma Z}$ are the cofactors of $g^{\mu\nu}$ in the $WW$ and $ZZ$ self-energies, respectively, and $s = \sin\theta_W$ and $c = \cos\theta_W$. The parameter $\Delta$ is often called the "universal" part of $\rho$, since, by definition, does not depend on the details of the process. According to the standard lore $\Delta$ contains the dominant contribution to $\rho$.

In this paper we address theoretical issues related to the definition and calculation of the universal part of the $\rho$ parameter. Several of the ingredients of the subsequent analysis are known, but they exist in a fragmented fashion and their relevance in the context of $\rho$-parameter calculations has not been fully recognized. The purpose of the present work is to provide a unified view of all relevant facts, incorporating them into a coherent framework.

The leading $m_t$ contributions to $\Delta$, both from one-loop and two-loop diagrams, are gauge-independent and ultraviolet finite However, as soon as the subleading contributions are considered, $\Delta$ becomes gauge-dependent and ultraviolet divergent. More specificaly, the leading one-loop $m_t$ contributions to $\Delta$ (of order $G_\mu m_t^2$) are manifestly gauge-independent, since the gauge fixing parameter does not appear inside fermion loops, and ultraviolet (UV) finite. On the other hand, as was explicitly shown by Degrassi, Kniehl, and Sirlin [2], the one-loop bosonic contributions to $\Delta$ (subleading in $m_t^2$, of order $g^2 m_t^0$) are *gauge dependent* and, except when formulated within a restricted class of gauges, UV divergent [3]. Similarly, the leading two-loop contributions to $\Delta$ [4-6](of order $G_\mu^2 m_t^4$) are also gauge-independent and UV-finite, exactly as their one-loop counterparts. On the other hand, the straighforward calculation of the subleading two-loop $m_t$ contributions (of order $G_\mu^2 m_t^2 M_Z^2$) to $\Delta$, carried out in the renormalizable Feynman gauge ($\xi_W = \xi_Z = 1$), gives rise to an answer which is ultraviolet divergent, in the sence that $\frac{1}{\epsilon}$ terms survive [7-8]. In addition, as stated in [7], the result for $\Delta$ is gauge dependent, in the context of the $R_\xi$ gauges. The reason for this rather striking analogy between the one-loop and the two-loop analysis, is the fact that, as soon as the subleading contributions (of order $m_t^0$ at one-loop, and of order $m_t^2$ at two-loops) are considered, Feynmam graphs containing the $W$ and $Z$ gauge-bosons must be included. It is the inclusion of such graphs, which, when carried out without a concrete guiding principle, gives rise to the aforementioned pathologies.

In order to understand the origin of the problems associated with the subleading contributions one has to first establish the mechanism which enforces the good behavior of the leading contributions, in particular their UV finiteness. If we denote the leading contributions (both at one and two loops) to the to $WW$ and $ZZ$ self-energies by $[\Pi_W^{\mu\nu}]^{(\ell)}$ and $[\Pi_Z^{\mu\nu}]^{(\ell)}$, respectively (the superscript $\ell$ stands for "leading"), and use the fact that

$A^{(\ell)}_{\gamma Z}(0) = 0$, we can write for $\Delta^{(\ell)}$:

$$\Delta^{(\ell)} = \frac{A^{(\ell)}_W(0)}{M^2_W} - \frac{A^{(\ell)}_Z(0)}{M^2_Z} \ . \tag{1.3}$$

It is elementary to show (see section 3) that $\Delta^{(\ell)}$ is finite as long as the following relation holds

$$\frac{1}{Z^{(\ell)}_W} = \frac{1}{Z^{(\ell)}_Z} + \left(\frac{\delta c^2}{c^2}\right)^{(\ell)} \tag{1.4}$$

where $Z_W$ and $Z_Z$ are the wave function renormalization constants of the $W$ and $Z$ fields, respectively, and

$$\left(\frac{\delta c^2}{c^2}\right)^{(\ell)} = \frac{Re[A^{(\ell)}_W(M^2_W)]}{M^2_W} - \frac{Re[A^{(\ell)}_Z(M^2_Z)]}{M^2_Z} \ . \tag{1.5}$$

Eq. (1.4) is indeed true for the leading contributions, due to a set of QED-like Ward identities relating the vertex and fermion wave function renormalization constants (the QED analog of $Z_1 = Z_2$). An equivalent way for establishing the finiteness of the leading contributions is to resort to the following Ward identities:

$$\begin{aligned} q_\mu q_\nu [\Pi^{\mu\nu}_W]^{(\ell)} &= M^2_W \Pi^{(\ell)}_\phi \\ q_\mu q_\nu [\Pi^{\mu\nu}_Z]^{(\ell)} &= M^2_Z \Pi^{(\ell)}_\chi \ , \end{aligned} \tag{1.6}$$

where $\Pi_\phi$ and $\Pi_\chi$ are the $\phi\phi$ and $\chi\chi$ self-energies, respectively, with $\phi$ ($\chi$) the charged (neutral) would-be Goldstone bosons. Writing the general self-energy in the form

$$\Pi_{\mu\nu}(q^2) = A(q^2) g_{\mu\nu} + B(q^2) \frac{q^\mu q^\nu}{q^2} \ , \tag{1.7}$$

and using the algebraic identity

$$A(0) = \left[\frac{d}{dq^2} \{q^\mu q^\nu \Pi_{\mu\nu}(q^2)\}\right]\bigg|_{q^2=0}\ , \tag{1.8}$$

together with the Ward identities of Eq. (1.6), and the fact that $A^{(\ell)}_{\gamma Z}(0) = 0$, we can write for $\Delta^{(\ell)}$:

$$\begin{aligned} \Delta^{(\ell)} &= \frac{A^{(\ell)}_W(0)}{M^2_W} - \frac{A^{(\ell)}_Z(0)}{M^2_Z} \\ &= \left[\frac{d}{dq^2} \{\Pi^{(\ell)}_\phi - \Pi^{(\ell)}_\chi\}\right]\bigg|_{q^2=0} \end{aligned}\ . \tag{1.9}$$

The final ingredient which enforces the finiteness of the result for the leading $\Delta\rho$ contributions, is the equality $Z_\phi^{(\ell)} = Z_\chi^{(\ell)}$, where $Z_\phi^{(\ell)}$ and $Z_\chi^{(\ell)}$ are the wave-function renormalization constants of $\Pi_\phi^{(\ell)}$ and $\Pi_\chi^{(\ell)}$, respectively (see section III). However, none of the above conditions are valid anymore, when one calculates the subleading parts of the $W$ and $Z$ self-energies in the framework of the $R_\xi$ gauges. Consequently, since the mechanism enforcing the finiteness does not operate any more, the resulting expressions do not have to be UV finite, and, indeed, they are not.

The standard way to circumvent the above problems is to readily abandon the notion of a "universal" part of $\Delta\rho$, by stating that, instead of only $\Delta$, the entire process must now be considered, in order to restore the finiteness and gauge-independence of the final answer. So, one has to introduce vertex and box corrections, which render the result gauge-idependent and finite, at the expense of making it *process-dependent*, and therefore non-"universal".

This unpleasant trade-off between gauge-independence and process-independence can be avoided however [2] if one defines the universal part of $\rho$ within the framework of the pinch technique (PT) [9-15]. As shown in Ref [2], the PT at one-loop gives rise to a gauge-independent and UV finite answer, *without* introducing any process-dependence. This is so because all PT self-energies are individually gauge-independent, and in addition, all conditions which enforce the finiteness are valid for *both* leading and subleading contributions. Evidently, the PT restores the mechanism for the cancellation of the UV divergences, and at the same time guarantees the gauge-and process-independence of the final answer.

In this paper we propose to elevate the $\Delta$ defined in terms of the PT self-energies as the truly universal part of $\Delta\rho$, beyond one-loop, in the spirit originally suggested in Ref [2]. This new quantity, which we denote by $\hat\Delta$, is endowed with three crucial properties:

i) $\hat\Delta$ is independent of the gauge-fixing procedure and the gauge-fixing parameter,

ii) $\hat\Delta$ is ultraviolet finite, and

iii) $\hat{\Delta}$ is process-independent.

In addition to the above obvious theoretical advantages, the calculation of $\hat{\Delta}$ is significantly fascilitated by the fact that only self-energy-like graphs need be considered. In particular, no vertex or box diagrams need be calculated, in order to render the answer finite, as is the case in the conventional treatment [16]. Therefore $\hat{\Delta}$ lends itself as the natural generalization of the conventional $\Delta$, which can consistently accomodate *both* leading and subleading contributions, and can be expressed in a closed analytic form up to two loops.

In Ref [7] the nature of the two-loop subleading contributions, as well as their numerical importance for the *conventionally* defined $\Delta$, were studied in the context of an $SU(2)$ model, with $M_W = M_Z = M$, and no photon. In Ref [8] the previous analysis was extended to the full standard model; the two answers turned out to be numerically rather close. Since we are mainly interested in addressing the conceptual issues involved, in this paper we also restrict ourselves to the study of this simplified version of the standard model. Of course, we have no a priori knowledge if the $\hat{\Delta}$ of the $SU(2)$ and the $\hat{\Delta}$ of the full stadard model will be numerically close, as was the case between the results of Ref [7] and Ref [8].

The paper is organized as follows: In section 2 we briefly review the PT, mainly as it applies in the present context. In section 3 we define the universal quantity $\hat{\Delta}$ and discuss some of its properties. In section 4 we calculate the subleading top quark contributions to $\hat{\Delta}$ in the framework of the $SU(2)$ model. This section is rather technical and contains several intermediate results. Finally, we present our conclusions in section 5. In addition, we present two Appendixes concerning scalar two loop integrals at zero momentum transfer, and the Feynman rules we have used.

## 2. The pinch technique

### Gauge-invariant effective self-energies

The simplest example that demonstrates how the PT works is the gluon two point function. [10] Consider the $S$-matrix element $T$ for the elastic scattering such as $q_1 \bar{q}_2 \to q_1 \bar{q}_2$, where $q_1, q_2$ are two on-shell test quarks with masses $m_1$ and $m_2$. To any order in perturbation theory $T$ is independent of the gauge fixing parameter $\xi$. On the other hand, as an explicit calculation shows, the conventionally defined proper self-energy depends on $\xi$ defined through the tree level gluon propagator

$$\Delta^{\mu\nu}(k,\xi) = \frac{-i}{k^2}[g^{\mu\nu} - (1-\xi)\frac{k^\mu k^\nu}{k^2}] \ . \tag{2.1}$$

At the one loop level this dependence is canceled by contributions from other graphs, which, at first glance, do not seem to be propagator-like. That this cancellation must occur and can be employed to define a g.i. self-energy, is evident from the decomposition:

$$T(s,t,m_1,m_2) = T_0(t,\xi) + T_1(t,m_1,\xi) + T_2(t,m_2,\xi) + T_3(s,t,m_1,m_2,\xi) \ , \tag{2.2}$$

where the function $T_0(t,\xi)$ depends kinematically only on the Mandelstam variable $t = -(\hat{p}_1 - p_1)^2 = -q^2$, and not on $s = (p_1 + p_2)^2$ or on the external masses. Typically, self-energy, vertex, and box diagrams contribute to $T_0$, $T_1$, $T_2$, and $T_3$, respectively. Such contributions are $\xi$ dependent, in general. However, as the sum $T(s,t,m_1,m_2)$ is g.i., it is easy to show that Eq. (2.2) can be recast in the form

$$T(s,t,m_1,m_2) = \hat{T}_0(t) + \hat{T}_1(t,m_1) + \hat{T}_2(t,m_2) + \hat{T}_3(s,t,m_1,m_2) \ , \tag{2.3}$$

where the $\hat{T}_i$ ($i = 0, 1, 2, 3$) are *individually* $\xi$-independent. The propagator-like parts of vertex and box graphs which enforce the gauge independence of $T_0(t)$, are called pinch parts. They emerge every time a gluon propagator or an elementary three-gluon vertex contributes a longitudinal $k_\mu$ to the original graph's numerator. The action of such a term is to trigger an elementary Ward identity of the form $\not{k} = (\not{p} + \not{k} - m) - (\not{p} - m)$ when it gets contracted with a $\gamma$ matrix. The first term removes (pinches out) the internal fermion propagator, whereas the second vanishes on shell. From the g.i. functions $\hat{T}_i$ ($i = 1, 2, 3$) one may now extract a g.i. effective gluon ($G$) self-energy $\hat{\Pi}_{\mu\nu}(q)$, g.i. $Gq_i\bar{q}_i$ vertices $\hat{\Gamma}^{(i)}_\mu$, and a g.i. box $\hat{B}$, in the following way:

$$\begin{aligned}
\hat{T}_0 &= g^2 \bar{u}_1 \gamma^\mu u_1 [(\frac{1}{q^2}) \hat{\Pi}_{\mu\nu}(q) (\frac{1}{q^2})] \bar{u}_2 \gamma^\nu u_2 , \\
\hat{T}_1 &= g^2 \bar{u}_1 \hat{\Gamma}^{(1)}_\nu u_1 (\frac{1}{q^2}) \bar{u}_2 \gamma^\nu u_2 , \\
\hat{T}_2 &= g^2 \bar{u}_1 \gamma^\mu u_1 (\frac{1}{q^2}) \bar{u}_2 \hat{\Gamma}^{(2)}_\nu u_2 , \\
\hat{T}_3 &= \hat{B} ,
\end{aligned} \quad (2.4)$$

where $u_i$ are the external spinors, and $g$ is the gauge coupling.

The one-loop expression for $\hat{\Pi}_{\mu\nu}(q)$, calculated in the Feynman gauge $\xi = 1$ is given by [10]:

$$\hat{\Pi}_{\mu\nu}(q) = \Pi^{(\xi=1)}_{\mu\nu}(q) + t_{\mu\nu} \Pi^P(q) , \qquad (2.5)$$

where $t_{\mu\nu} = -g_{\mu\nu} + q^\mu q^\nu/q^2$, $l_{\mu\nu} = q^\mu q^\nu/q^2$ are the usual transverse and longitudinal projectors, and

$$\Pi^P(q) = 2i c_a g^2 q^2 \int_D \frac{1}{k^2(k+q)^2} , \qquad (2.6)$$

where $\int_D \equiv \mu^{4-D} \int \frac{d^D k}{(2\pi)^D}$ is the dimensionally regularized loop integral, $D$ is the dimensionality of space-time, and $c_a$ is the quadratic Casimir operator for the adjoint representation [for $SU(N)$, $c_a = N$]. After integration and renormalization we find

$$\Pi^P(q) = 2 c_a (\frac{g^2}{16\pi^2}) q^2 \ln(\frac{-q^2}{\mu^2})] . \qquad (2.7)$$

Adding this to the Feynman-gauge proper self-energy

$$\Pi^{(\xi=1)}_{\mu\nu}(q) = [\frac{5}{3} c_a (\frac{g^2}{16\pi^2}) q^2 \ln(\frac{-q^2}{\mu^2})] t_{\mu\nu} , \qquad (2.8)$$

we obtain for $\hat{\Pi}_{\mu\nu}(q)$

$$\hat{\Pi}_{\mu\nu}(q) = b g^2 \ln(\frac{-q^2}{\mu^2}) t_{\mu\nu} , \qquad (2.9)$$

where $b = \frac{1}{16\pi^2}\frac{11c_a}{3}$ is the coefficient of $-g^3$ in the usual $\beta$ function of QCD without fermions. This procedure can be extended to an arbitrary $n$-point function; of particular physical interest are the g.i. three and four point functions $\hat{\Gamma}_{\mu\nu\alpha}$ [17-18] and $\hat{\Gamma}_{\mu\nu\alpha\beta}$ [19]. Finally, the generalization of the PT to the case of non-conserved external currents is technically more involved, but conceptually straightforward [20-21].

## The current algebra formulation of the pinch technique

An important alternative formulation of the PT in the context of the SM has been introduced by Degrassi and Sirlin Ref. [12]. In this approach the interaction of gauge bosons with external fermions is expressed in terms of current correlation functions, i.e. matrix elements of Fourier transforms of time-ordered products of current operators [22]. This is particularly economical because these amplitudes automatically include several closely related Feynman diagrams. When one of the current operators is contracted with its four-momentum (i.e. the four momentum absorbed by the current), a Ward identity is triggered. The pinch part is then identified with the contributions involving equal-time commutators in the Ward identities, and therefore involve amplitudes in which the number of current operators has been decreased by one or more. As emphasized in Ref. [12], this procedure has an important advantage when one considers external particles endowed with strong interactions. Because the contributions from the equal-time commutators are not affected by the dynamics of the strong interactions, the aforementioned identification ensures the universality of the "pinch parts". That is, the cofactors of the current operators in the pinch parts are the same whether the external particles are leptons or strongly interacting fermions. To illustrate the method with an example, consider the vertex function $iU_z^\lambda(W)$ that contributes to Fig.(1b), where now the gauge particles in the loop are W's, the incoming one is a $Z$, and the incoming and outgoing fermions are massless. It can be

written as

$$iU_z^\lambda(W) = \frac{ig^3c}{2} \int \frac{d^nk}{(2\pi)^n} \Delta_{\alpha\rho}^W(k)\Delta_{\sigma\beta}^W(k+q)[g^{\rho\sigma}(2k+q)^\lambda - g^{\lambda\sigma}(2q+k)^\rho - g^{\rho\lambda}(k-q)^\sigma]$$
$$\times \int d^n x e^{ikx} < f|T^*[J_W^{\alpha\dagger}(x)J_W^\beta(0)]|i >, \tag{2.10}$$

where

$$\Delta_i^{\mu\nu}(k,\xi_i) = \frac{-i}{k^2 - M^2}[g^{\mu\nu} - (1-\xi_i)\frac{k^\mu k^\nu}{k^2 - \xi_i M_i^2}] \tag{2.11}$$

with $i = W, Z, \gamma$ and $M_\gamma = 0$, are the propagators of the gauge bosons in a general $R_\xi$ gauge. An appropriate momentum, say $k_\alpha$, from the three gauge boson vertex or the longitudinal part of the propagator can be transformed into a derivative $\frac{d}{dx_\alpha}$ acting on the $T^*$ product. Invoking current conservation this leads to an equal-time commutator of current operators. Thus, such contribution are proportional to the matrix element of a single current operator, namely $< f|J_3^\lambda|i >$; these are precisely the pinch parts. Calling $iU_z^\lambda(W)_P$ the total pinch contribution from Eq. (2.10), we find in the $\xi = 1$ gauge

$$U_z^\lambda(W)_P = ig^3c < f|J_3^\lambda|i > \int \frac{1}{(k^2 - M_w^2)[(k+q)^2 - M_w^2]}. \tag{2.12}$$

Clearly, the integral in Eq. (2.12) is the generalization of the QCD expression Eq. (2.6) to the massive gauge boson case.

### Ward identities of the PT

Another important fact is that the PT Green's functions satisfy *tree-level* Ward identities. Most noticeably, the g.i. QCD vertex $Gq_i\bar{q}_i$ satisfies the following Ward-identity:

$$q^\mu \hat{\Gamma}_\mu = \hat{\Sigma}(p+q) - \hat{\Sigma}(p) \tag{2.13}$$

where $\hat{\Sigma}(p)$ is the g.i. quark self-energy [23] The above QED-like Ward identity, which is not true for the conventional $\Gamma_\mu^{(i)}$, enforces the equality $\hat{Z}_1 = \hat{Z}_2$ between the vertex renormalization constant $\hat{Z}_1$ and the quark wave function renormalization constant $\hat{Z}_2$. Consequently, exactly as happens in QED, the PT vacuum polarization contains the entire running of the QCD coupling, as shown already by the explicit result of Eq. (2.9) [10], [24].

The above QCD results have been generalized for the electroweak part of the standard model, where it was found that the one-loop PT Green's functions satisfy again tree-level Ward identities [18], [21], [25]. Therefore, the wave function renormalizations for the PT $\gamma\gamma$ and $WW$ self-energy contain the running of the gauge couplings $e^2(q^2)$ and $g^2(q^2)$, respectively [12], [13]. Denoting by $\widehat{\Pi}_{\mu\nu}$ the gauge boson PT self energies ($\widehat{\Pi}^W_{\mu\nu}$ or $\widehat{\Pi}^Z_{\mu\nu}$), by $\widehat{\Theta}_\mu$ the mixed PT self energy of a gauge boson and its associated unphysical scalar ($\widehat{\Pi}^{W^-\phi^+}_\mu \equiv \widehat{\Pi}^+_\mu = -\widehat{\Pi}^-_\mu$, or $\widehat{\Pi}^{Z\chi}_\mu$), and by $\widehat{\Omega}$ the PT self energies of the unphysical scalars ($\widehat{\Pi}^\phi$ or $\widehat{\Pi}^\chi$), the following WI hold [26]:

$$q^\mu \widehat{\Pi}_{\mu\nu} - iM\widehat{\Theta}_\nu = 0$$
$$q^\mu \widehat{\Theta}_\mu + iM\widehat{\Omega} = 0 \qquad (2.14)$$
$$q^\mu q^\nu \widehat{\Pi}_{\mu\nu} - M^2 \widehat{\Omega} = 0 \ .$$

Additional WI between other PT Green's functions can be found in the literature. As was explained in detail in [21], the PT Ward identities are instrumental for the final cancellation of gauge dependences in S-matrix elements.

Imposing the elementary requirement that the renormalized PT Green's functions should respect the same Ward identities as their unrenormalized counterparts, we obtain the following relationships between the standard model renormalization constants:

$$\widehat{Z}_W = \widehat{Z}_g^{-2} \qquad (2.15)$$

$$\widehat{Z}_Z^{-1} = \widehat{Z}_W^{-1} + \frac{\delta \hat{c}^2}{\hat{c}^2} \ , \qquad (2.16)$$

and

$$\widehat{Z}_H = \widehat{Z}_\chi = \widehat{Z}_\phi = \widehat{Z}_W + \frac{\delta \hat{M}_W^2}{\hat{M}_W^2} \ , \qquad (2.17)$$

with

$$\frac{\delta \hat{c}^2}{\hat{c}^2} = \frac{\delta \hat{M}_W^2}{\hat{M}_W^2} - \frac{\delta \hat{M}_Z^2}{\hat{M}_Z^2} \ . \qquad (2.18)$$

In the simplified $SU(2)$ model we will use laetr we have:

$$\widehat{Z}_Z = \widehat{Z}_W = \widehat{Z}_g^{-2} \tag{2.19}$$

and

$$\widehat{Z}_H = \widehat{Z}_\chi = \widehat{Z}_\phi = \widehat{Z}_g^{-2} + \frac{\delta \widehat{M}^2}{\widehat{M}^2} \ . \tag{2.20}$$

As usual, gauge boson self energies are cast in the form

$$\widehat{\Pi}_{\mu\nu}(q) = g_{\mu\nu} \hat{A}(q^2) + \frac{q_\mu q_\nu}{q^2} \hat{B}(q^2) \tag{2.21}$$

and the renormalization constants are defined from the expansion

$$\hat{A}(q^2) = \hat{A}(M^2) + (q^2 - M^2) \frac{d\hat{A}(q^2)}{dq^2}\bigg|_{q^2 = M^2} + \hat{A}(q^2)^{finite} \tag{2.22}$$

as

$$\delta \hat{M}^2 = Re\left(\hat{A}(M^2)\right) \ , \qquad \widehat{Z}^{-1} = 1 - \frac{d\hat{A}(q^2)}{dq^2}\bigg|_{q^2 = M^2} \ . \tag{2.23}$$

## 3. The universal $\hat{\Delta}$

In this section we focus on the universal part of $\rho$ defined by the PT. In particular we will emphasize issues of gauge-independence and finiteness.

Traditionally the universal part $\Delta$ is defined as in Eq. (1.2). $\Delta$ vanishes in the limit of exact $SU(2)_V$ custodial symmetry, e.g. for $M_W = M_Z$ (no hypercharge) and for degenerate fermion doublets, $m_u = m_d$ [27].

The fermionic one-loop contribution is given by [1]

$$\Delta\rho_f^1 = N_c \frac{G_\mu}{8\sqrt{2}\ \pi^2} \left[ m_u^2 + m_d^2 + \frac{2m_u^2 m_d^2}{m_u^2 - m_d^2} \ln \frac{m_d^2}{m_u^2} \right] \tag{3.1}$$

Clearly, $\Delta\rho^1 \to 0$ as $m_u \to m_d$. When the mass splitting in the fermion doublet is large, as in the case of the top and bottom quarks, the factor in square brackets in Eq. (3.1) is replaced by $m_t^2$, the heavy fermion mass. By neglecting the contribution of all light fermions the one loop fermionic contribution is written as

$$\Delta\rho_f^1 = N_c x_t \tag{3.2}$$

where

$$x_t = \frac{G_\mu m_t^2}{8\sqrt{2}\,\pi^2} = \frac{g^2}{16\pi^2}\frac{m_t^2}{4M_W^2} \tag{3.3}$$

If one attempts to use the definition of Eq. (1.2) to include bosonic one loop corrections one is faced with two problems

(a) The result is $\xi - dependent$

(b) Unless computed in a special class of gauges, it is ultraviolet divergent.

In particular, regarding the first point, the dependence on the gauge fixing parameter enters through the tree-level propagators for the $W$, the $Z$ and the photon which, in the $R_\xi$ gauges they are given by Eq. (2.11). In addition, the tree-level propagators of the unphysical Goldstone bosons are given by

$$\Delta_s(q, \xi_i) = \frac{i}{q^2 - \xi_i M_i^2} \quad , \tag{3.4}$$

with $(s, i) = (\phi, W)$ or $(\chi, Z)$, and they also explicitly depend on $\xi_i$. The conventional one-loop self-energies depend explicitly on the gauge-fixing parameters, even at $q^2 = 0$.

Regarding point (b), unless the relation

$$\xi_W = \xi_\gamma sin^2\theta + \xi_Z cos^2\theta \ . \tag{3.5}$$

between the gauge fixing parameters $\xi_i$ is satisfied, the resulting expression for $\Delta$ contains a term proportional to $\frac{1}{\epsilon}$.

The problems mentioned above persist when one computes the two-loop contributions to $\Delta$. Again, as happens in the one-loop case, the leading contributions are both gauge-independent and finite. As soon as the subleading contributions are taken into account the pathologies familiar from the one-loop, reappear: The results are again gauge-dependent and, even when computed in the Feynman gauges $\xi_W = \xi_Z = \xi_\gamma = 1$, which obviously satisfies the one-loop condition of Eq. (3.5), are ultraviolet divergent. Evidently, Eq. (3.5) breaks down beyond one loop.

Before we proceed to the study of $\Delta$ defined via the PT, it is worthwhile to further elaborate on the gauge independence of the leading two-loop contributions to the conventional $\Delta$ mentioned above. The reason for it is simply that the S-matrix is gauge-independent, and there are no vertex or box contributions proportional to $m_t^4$ which could cancell any possible gauge-dependences coming from the self-energy graphs. Consequently, one is allowed to choose any convenient gauge for calculating these leading self-energy contributions. In particular, in the Feynman gauge ($\xi = 1$)only graphs with scalars and fermions contribute to this order. This is not generally true however for an arbitrary value of the gauge fixing parameter $\xi$. The graphs of Fig.9 for example will give rise to leading $m_t^4$ contributions, due to the longitudinal parts of the gauge boson propagators. Obviously the characterization of individual Feynman graphs as "leading" and "subleading" is a *gauge dependent* statement. When all relevant graphs are correctly accounted for, they will indeed conspire to furnish a unique $\xi$-independent answer, which will clearly be identical to the one obtained in the Feynman gauge. It is instructive to briefly highlight the mechanism enforcing the cancellations of the gauge-dependences. To that end we can employ the elementary algebraic identity

$$\frac{1}{q^2 - \xi M^2} = \frac{1}{q^2 - M^2} + \frac{(1-\xi)M^2}{(q^2 - M^2)(q^2 - \xi M^2)} \tag{3.6}$$

in the graphs containing scalars and fermions. The first term in r.h.s. of Eq. (3.6) is the Feynam gauge scalar propagator, whereas the second term resembles the longitudinal part of the corresponding gauge boson propagator. The final cancellation proceeds after using the elementary Ward identity

$$k_\mu \gamma^\mu P_L \equiv \slashed{k} P_L = S_i^{-1}(p+k)P_L - P_R S_j^{-1}(p) + m_i P_L - m_j P_R \ , \tag{3.7}$$

where $P_{R,L} = (1 \pm \gamma_5)/2$, triggered by the longitudinal term $k^\mu k^\nu$ of the gauge boson propagator. It is this Ward identity, which when applied for both $k^\mu$ and $k^\nu$, gives rise to an extra power of $m_t^2$, thus converting pieces of a diagram, which is subleading in the Feynman gauge, into leading.

If one defines $\Delta$ instead in terms of the effective $WW$ and $ZZ$ propagators obtained via the PT all problems associated with the gauge-independence and finiteness of the subleading parts are automatically solved. We denote $\hat{\Delta}$ the universal part of $\Delta\rho$ defined via the PT as follows

$$\hat{\Delta} = \frac{\hat{A}_W(0)}{M_W^2} - \frac{\hat{A}_Z(0)}{M_Z^2} \tag{3.8}$$

It is important to notice the absence of the $\gamma Z$ mixing term in the above definition; this is so because in the PT the $\gamma Z$ self-energy vanishes at $q^2 = 0$, e.g. $\hat{\Pi}_{\mu\nu}^{\gamma Z}(0) = 0$. The PT self-energies are individually independent of the gauge-fixing parameters, and when combined according to Eq. (3.8) they give a UV finite answer.

Although the gauge-invariance of the result in the context of the PT is guaranteed by construction, its finiteness may be less obvious. There are two equivalent ways of understanding why the PT definition gives rise to a finite expression, both relying on the Ward identities presented in section 2. Writing the $WW$ and $ZZ$ self-energies in the form

$$\hat{A}_W(q^2) = \hat{A}_W(M_W^2) + (q^2 - M_W^2)[1 - \hat{Z}_W^{-1}] + \hat{A}_W^f(q^2) \; , \tag{3.9}$$

and

$$\hat{A}_Z(q^2) = \hat{A}_Z(M_Z^2) + (q^2 - M_Z^2)[1 - \hat{Z}_Z^{-1}] + \hat{A}_Z^f(q^2) \; , \tag{3.10}$$

Eq. (3.8) yields

$$\hat{\Delta} = \hat{\Delta}\big|_{div} + \frac{\hat{A}_W^f(0)}{M_W^2} - \frac{\hat{A}_Z^f(0)}{M_Z^2} \tag{3.11}$$

$\hat{\Delta}\big|_{div}$, which contains the terms proportional to $\frac{1}{\epsilon}$ ( and possibly finite pieces, which we neglect at this point), is given by

$$\begin{aligned}\hat{\Delta}\big|_{div} &= \left[\frac{\hat{A}_W(M_W^2)}{M_W^2} - \frac{\hat{A}_Z(M_Z^2)}{M_Z^2}\right] + (\hat{Z}_W^{-1} - \hat{Z}_Z^{-1}) \\ &= 0\end{aligned} \tag{3.12}$$

where in the last step we used Eq. (2.16). So, $\hat{\Delta}$ of Eq. (3.11) is finite for leading, subleading, and bosonic contributions.

Another way to establish the finiteness of $\hat{\Delta}$ is the following: After computing the off-shell $WW$ and $ZZ$ self-energies, which have the form of Eq. (1.7), we use Eq. (1.8) together with the Ward identities of Eq. (1.6), we can write for $\hat{\Delta}$:

$$\begin{aligned}\hat{\Delta} &= \frac{\hat{A}_W(0)}{M_W^2} - \frac{\hat{A}_Z(0)}{M_Z^2} \\ &= [\frac{d}{dq^2}\{\hat{\Pi}_\phi - \Pi_\chi\}]|_{q^2=0}\end{aligned} \quad (3.13)$$

We than use the fact that

$$\Pi_\phi(q^2) = \Pi_\phi(M_W^2) + (q^2 - M_W^2)[1 - \widehat{Z}_\phi^{-1}] + \Pi_\phi^f(q^2)$$

and

$$\Pi_\chi(q^2) = \Pi_\chi(M_Z^2) + (q^2 - M_Z^2)[1 - \widehat{Z}_\chi^{-1}] + \Pi_\chi^f(q^2)$$

where $1 - \widehat{Z}_\phi^{-1} = \frac{d\widehat{\Pi}_\phi(q^2)}{dq^2}|_{q^2=M_W^2}$ and $1 - \widehat{Z}_\chi^{-1} = \frac{d\widehat{\Pi}_\chi(q^2)}{dq^2}|_{q^2=M_Z^2}$. Taking the difference of the two scalar self-energies, and using the fact that $\widehat{Z}_\phi = \widehat{Z}_\chi \equiv \widehat{Z}$ we have

$$\widehat{\Pi}_\phi(q^2) - \widehat{\Pi}_\chi(q^2) = [\widehat{\Pi}_\phi(M_W^2) - \widehat{\Pi}_\chi(M_Z^2)] + (1 - \widehat{Z}^{-1})[M_Z^2 - M_W^2] + [\widehat{\Pi}_\phi^f(q^2) - \widehat{\Pi}_\chi^f(q^2)] \ . \quad (3.14)$$

The first two terms on r.h.s. of the last equation are proportional to $\frac{1}{\epsilon}$, but they are constant, independent of $q^2$. Therefore, upon differentiation with respect to $q^2$ they vanish. The third term is $q^2$-dependent and finite, and after diferentiating it with respect to $q^2$ and subsequently setting $q^2 = 0$ we obtain the UV finite expression for $\hat{\Delta}$. Clearly, the above proof of finiteness does not depend on the choice of the renormalization point; so instead of expanding around $q^2 = M_W^2$ and $q^2 = M_Z^2$, we can equally well expand around $q^2 = \mu_1^2$ and $q^2 = \mu_2^2$, respectively.

It is important to emphasize that all properties of the PT self-energies stemming from the PT Ward identities hold for the corresponding conventional self-energies computed in the background field method (BFM) [28], for *every* value of the gauge-fixing parameter $\xi_Q^W$ and $\xi_Q^Z$, used for the quantum fields, and to *all orders* in perturbation theory [29]. Consequently, in the BFM the finiteness of $\Delta$ is true for every value of the gauge fixing parameter and to all loops [30]. The final answer is however *not* gauge-invariant. This is so because in the BFM the gauge boson self-energies depend in general on the choice of gauge-fixing parameters [23](these gauge dependent terms are however UV finite); this remaining gauge-dependence does *not* cancel when the difference of the $WW$ and $ZZ$ self-energies is formed, in order to construct $\Delta$, already at the one-loop level. So, the one loop bosonic contribution $\Delta_b^1\big|_{BFM}$, defined via the BFM *off shell* $\phi\phi$ and $\chi\chi$, which are both explicitly $\xi_Q$-dependent, has the form

$$\Delta_b^1\big|_{BFM} = \Delta_b^1\big|_{\xi_Q=1} + \Delta_b^1(\xi_Q). \tag{3.15}$$

$\Delta\big|_{\xi_Q=1} = \hat{\Delta}_b^1$ is the gauge independent bosonic PT result, given by

$$\hat{\Delta}_b^1 = \frac{g^2}{64\pi^2}\left[1 + \frac{9c^2 - 8c^4 + h(5c^2 - 6)}{s^2(c^2 - h)}\log(c^2) - \frac{3s^2 h^2}{c^2(c^2 - h)(1 - h)}\log(h)\right], \tag{3.16}$$

where $h = M_H^2/m_t^2$, and $M_H$ is the mass of the Higgs scalar. $\Delta_b^1(\xi_Q)$ carries explicitly the gauge parameter dependence (one sets for simplicity $\xi_Q^Z = \xi_Q^W \equiv \xi_Q$). $\Delta_b^1(\xi_Q)$ vanishes at $\xi_Q = 1$, and when $M_W = M_Z$ ($s = 0$), but is non-zero otherwise; its explicit expression has been reported in [31]. As one can see from Fig.1, the gauge dependent term $\Delta_b^1(\xi_Q)$ is unbounded from above and below, and is numerically significant. We observe that although the BFM endows the Green's function with the desirable theoretical properties, it fails to address the crucial issue of gauge-fixing parameter independence, as any other gauge fixing procedure for that matter. Nevertheless, it provides a convenient starting point for the implementation of the PT [23].

# 4. Subleading top contributions to $\hat{\Delta}$ in the $SU(2)$ model

In this section we apply the formalism developed thus far to the case of an $SU(2)$ model, which corresponds to the standard $SU(2) \times U(1)$, with electromagnetism turned off. This means that $s = 0$ and $c = 1$, or equivalently $M_W = M_Z \equiv M$, and there is no photon. In such a model, a non-vanishing $\hat{\Delta}$ comes only from the mass-splitting within a fermion iso-doublet; in particular, there are no genuinely bosonic contributions to $\hat{\Delta}$, since $M_W = M_Z$. We ignore the contributions of all light fermions, and concentrate on the top quark contribution. Since at one loop $\hat{\Delta}$ contains only fermionic contributions it is automatically gauge independent and UV finite. Obviously, at one loop the PT definition coincides with the conventional one.

At two loops there are two kinds of contributions:

(i) The leading, of order $m_t^4$, which originate from graphs containing fermions and only scalars, without gauge bosons.

(ii) The sub-leading, of order $m_t^2$, which originate from the graphs of (i), if scalars are replaced by gauge bosons ($W$ or $Z$).

The leading contributions of a very heavy top quark to the conventionally defined $\Delta$ have been first computed in [4], in the limit where $M_W = M_Z = M_H = 0$. The case of an arbitrary Higgs mass $M_H$, but still $M_W = M_Z = 0$, was computed in [5], [6], and the case $M_W = M_Z = M \neq 0$ was presented in [7]. In the above calculations the Feynman gauge was used; as already explained in section 3, this convenient choice of gauge is legitimate, since the result is guaranteed to be gauge-independent and UV finite. Clearly, the PT and the conventional definitions are identical for the leading part of the calculation.

The sub-leading top contributions to the conventionally defined $\Delta$ were first addressed in [7]; it was explicitly shown that the resulting expressions contain left-over terms proportional to $\frac{1}{\epsilon}$. In addition, it was pointed out that this result, calculated in the renormalizable Feynman gauge, was in fact gauge-dependent, and was correctly argued that due to this theoretical shortcommings the inclusion of subleading corrections deprives the conventionally defined $\Delta$ of any physical meaning. In order to restore gauge invariance and finiteness several contributions from vertex and box diagrams were included; however, since no guiding principle such as the PT was followed, these contributions rendered the answer process-dependent. Furthermore, it was sugested that, since the sub-leading contributions cannot be defined in a process-independent way, the possibility of resumming them [32] should probably be abadoned. Since no closed expressions for two-loop vertex and box graphs exist yet in the literature, the process-dependent parts were calculated approximately, up to order $\mathcal{O}(M^2/m_t^2)$, for the case of $\nu_\mu e$ scattering. The final conclussion was that the part of $\delta\rho$ that one extracts with their method, for the case of $\nu_\mu e$ scattering receives sizable corrections due to sub-leading top contributions.

In the context of the PT all aforementioned pathologies are automatically bypassed. The answer is gauge-independent by construction, UV finite, and manifestly process-independent. We conclude therefore that in the context of the PT there is no limitation whatsoever in defining the subleading top contributions to the universal $\hat{\Delta}$. In particular, the necessary condition for attempting the resummation of the universal part of $\rho$, i.e. the process-idependence, is still valid. It turns out that the relative size of the subleading contributions compared to the leading is in accordance to what one would naively expect from a power series whose expansion parameter $r = M^2/m_t^2$ is of the order of 1/4.

We now proceed to the more technical aspects of our calculations.

(a) It has been known for years that when computing the PT Green's functions any convenient gauge may be chosen, as long as one properly accounts for the pinch contributions within that gauge [10]. In the context of the "renormalizable" $R_\xi$ gauges the most convenient gauge-fixing choice is the Feynman gauge ($\xi = 1$). This is so because the longitudinal parts of the gauge boson propagators, which can pinch, vanish for $\xi = 1$, and the only possibility for pinching stems from the tree-boson vertices. As was recently realized [29], the task of the PT rearrangment of the S-matrix can be further fascilitated, if one quantizes the theory in the context of the BFM. Even though the Feynman rules obtained via the BFM are rather involved, they become particularly suited for one-loop pinching, if one chooses the Feynman gauge ($\xi_Q = 1$) inside the quantum loops. In fact, all possible one-loop pinch contribution are zero in this gauge; consequently, the one-loop PT Green's functions (which one can obtain for *every* gauge) are *identical* to the *conventional* Green's functions, calculated in the Feynman gauge of the BFM. This property of the Feynman gauge in the BFM persists in two loop calculations, *only* for the subset of diagrams which contain at least one fermion-loop [33], which is precisely the type of graphs we are interested in [34]. We therefore choose to work in the BFM Feynman gauge. This correspondence between PT and BFM at $\xi_Q = 1$ [35] breaks down for the two-loop purely bosonic part [36]. The technical details leading to this conclusion will be presented in [37].

(b) Using the algebraic identity of Eq. (1.8), and the WI of Eq. (2.14), we write $\hat{\Delta}$ as in Eq. (3.13). Therefore, the entire calculation boils down to calculating the derivative of each Feynman graph shown in Fig.2-Fig.7, at $q^2 = 0$. In these figures we show the complete set of two-loop irreducible graphs that contribute only $m_t^2$ terms to $\hat{\Delta}$, in the $SU(2)$ model. Graphs with photons or with couplings proportional to $\sin\theta_W$ in the standard model are omitted. For example, graphs such as those shown in Fig.8 for the self-energies of the scalars do not contribute, since they contain couplings proportional to $\sin\theta_W = 0$. In our calculation we have used a fully anticommuting $\gamma_5$ since this does not produce any inconsistencies for the graphs we have to compute.

The validity of the Ward identity of Eq. (2.14) for the full standard model has been verified by directly contracting individual graphs by $q^\mu q^\nu$, *before* carrying out any loop integrations. In fact, the Ward identities of Eq. (2.14) hold individually for each of the graphs shown in Fig.2- Fig.7, where the corresponding graph of the gauge boson self energy –which is to be contracted by $q^\mu q^\nu$ – can be obtained by replacing the external $\hat{\phi}\hat{\phi}$ ( $\hat{\chi}\hat{\chi}$) legs by $\hat{W}\hat{W}$ ( $\hat{Z}\hat{Z}$), respectively. The only exceptions are some of the triangle graphs, i.e.

Fig.5(a,c), that need to be combined with graphs such as those depicted in Fig.8, in order to yield zero in the r.h.s. of the WI. For example we have

$$q^\mu q^\nu \left[\Pi_{\mu\nu}^{W\ (5.a)} + \Pi_{\mu\nu}^{W\ (8.c)}\right] - M_W^2 \left[\Pi^{\phi\ (5.a)} + \Pi^{\phi\ (8.d)}\right] = 0 \tag{4.1}$$

$$q^\mu q^\nu \left[\Pi_{\mu\nu}^{W\ (5.c)} + \Pi_{\mu\nu}^{W\ (8.a)}\right] - M_W^2 \left[\Pi^{\phi\ (5.c)} + \Pi^{\phi\ (8.b)}\right] = 0 \tag{4.2}$$

In the above equations, we note that although the graphs of Fig.8 for the self energies of the scalars vanish for $\sin\theta_W = 0$, the corresponding ones for the gauge bosons do not.

Using the notation

$$G_k(q^2) \equiv -i \ (\text{Feynman} - \text{Graph } k)$$

we find it more convenient to act with the four-Laplacian instead of the regular derivative of $q^2$. Namely

$$\left.\frac{dG_k(q^2)}{dq^2}\right|_{q^2=0} \equiv \frac{1}{2D}\left(\Box_q G_k(q^2)\right)_{q=0}. \tag{4.3}$$

This facilitates the computation enormously since it reduces it to straightforward algebra that can be carried out easily by hand. This procedure reduces each graph down to standard scalar two loop integrals at zero external momentum, for which explicit expressions can easily be found (see Appendix I). As it turns out the result can be analyticaly expressed, like the leading corrections, in terms of logarithms, dilogarithms and the two functions $g(x)$ and $f(x, 1)$. The function $g(x)$ stems from the on-shell counterterms, while $f(x, 1)$ from the two loop scalar integrals; they too can be expressed in terms of logarithms and dilogarithms. Explicit intermediate results for each of the set of graphs of Fig.2-Fig.7 are given in the next section.

# 5. Calculations and results

In this section we present the explicit results of the two loop one particle irreducible contributions $\hat{\Delta}^{(2)}$ to the universal part $\hat{\Delta}$ of the $\rho$ parameter. We neglect the contributions of the light fermions and consider only the effect of a single fermion doublet $(t, b)$, with large mass splitting. The mass of the lightest partner in the doublet has been set equal to zero ($m_b = 0$) from the beginning of the calculation; this has produced no mass singularities. We decompose $\Delta^{(2)}$ in three parts as :

$$\hat{\Delta}^{(2)} = \hat{\Delta}^{(2)}_{lead} + \hat{\Delta}^{(2)}_{sub} + \hat{\Delta}^{(2)}_{bos} \tag{5.1}$$

where $\hat{\Delta}^{(2)}_{lead}$ are the graphs with scalars and fermions that contain leading contributions proportional to $m_t^4$ as well as $m_t^2$, $\hat{\Delta}^{(2)}_{sub}$ are the graphs with fermions and gauge bosons that contain contributions of order $m_t^2$ only, and finally $\hat{\Delta}^{(2)}_{bos}$ is the pure bosonic contribution which is independent of $m_t$ and in our approximation is zero. The results are given in terms of two variables $h = M_H^2/m_t^2$, and $r = M^2/m_t^2$, where $M_H$ is the mass of the Higgs boson and $m_t$ is the pole mass of the heavy quark (top) in the doublet.

The leading contributions $\hat{\Delta}^{(2)}_{lead}$ are given by [38]

$$\begin{aligned}
\hat{\Delta}^{(2)}_{lead} =&\ N_c x_t^2\ R(h, r) \\
=&\ N_c x_t^2 \left\{ 23 - 4h + \frac{2h}{h-r} - 11r + \frac{\pi^2}{3} r^2 - \left(2 - \frac{h}{2}\right) \sqrt{h} g(h) + \frac{r}{2} \sqrt{r} g(r) + r^2 \ln^2 r \right. \\
&- (1-r)^2 \ln(1-r) + \left[ -6 - 2h - 12r + \frac{3}{2} r^2 + \frac{2h^2}{(h-r)^2} + 2h \frac{(1+h)}{h-r} + \frac{8}{4-r} \right] \ln r \\
&+ \left[ -4h + \frac{h^2}{2} - \frac{2h^2}{(h-r)^2} - 2h \frac{(2+h)}{h-r} \right] \ln h \\
&+ \left[ -(1-h)^2 + 2h \frac{(1-h)^2}{(h-r)^2} + 2 \frac{(1-h)^3 + (1-h)^2}{h-r} \right] \text{Li}_2(1-h) \\
&+ \left[ -11 + 10h - 2h^2 - 2hr + 16r - 5r^2 - 2h \frac{(1-h)^2}{(h-r)^2} - 2 \frac{(1-h)^3 + (1-h)^2}{h-r} \right] \text{Li}_2(1-r)
\end{aligned}$$

$$+ \left[ 6 - 4h + h^2 + 2h^2 \frac{(4-h)}{(h-r)^2} - 2h \frac{1-(3-h)^2}{h-r} \right] f(h,1)$$

$$+ \left[ 26 - 14h + 2h^2 + 2hr - 30r + 7r^2 - 2h^2 \frac{(4-h)}{(h-r)^2} - 2h \frac{1+(3-h)^2}{h-r} + \frac{4}{4-r} \right] f(r,1) \Bigg\}$$

(5.2)

where $f(x,1)$ is a function stemming from the two-loop scalar integrals [39-40], [6]

$$f(x,1) = -\frac{1}{\sqrt{4-y}} \left[ \frac{\pi^2}{6} + 2\text{Li}_2(\zeta) + \frac{1}{2} \ln^2 \zeta \right] \qquad 4 \leq x$$

$$= -4 \ln 2 \qquad x = 4 \qquad (5.3)$$

$$= -\frac{2}{\sqrt{y-4}} \text{Cl}_2(\varphi) \qquad 0 \leq x \leq 4 \ ,$$

and $g(x)$ is a function that originates from the on-shell counterterms

$$g(x) = \sqrt{4-x}\,[\pi - \phi] \qquad 0 \leq x \leq 4$$

$$= 0 \qquad x = 4 \qquad (5.4)$$

$$= \sqrt{x-4} \ln(-\zeta) \qquad 4 \geq x \ .$$

The variables $y$, $\zeta$, and $\phi$ are defined as in [6]

$$y = \frac{4}{x} \ , \qquad \zeta = \frac{\sqrt{1-y}-1}{\sqrt{1-y}+1} \ , \quad \text{and} \ \phi = 2 \ \arcsin\left(\frac{\sqrt{x}}{2}\right) \ .$$

By taking the $r \to 0$ limit in Eq. (5.2) one can recover the result of Eq.(12) of [6] where the leading contributions were calculated in the approximation $M_Z = M_W = 0$. In the analytic formula of Eq. (5.2) the apparent mass singularities cancell in the relevant limits $h \to r$ and $r \to 4$. For example, in the $r \to 4$ limit the terms that contain $r - 4$ in their denominator give $4(2 \ln r + f(r,1))/(4-r) \to (4/3)(-1+\ln 2)$. In the limit $h \to r$ Eq. (5.2) reduces to

$$\hat{\Delta}_{lead}^{(2)}\bigg|_{h=r} = N_c x_t^2 \bigg\{ 19 - 17r + \frac{\pi^2}{3} r^2 - (2-r)\sqrt{r}g(r) + r^2 \ln^2 r - (1-r)^2 \ln(1-r)$$

$$- 2r(10-r) \ln r + 2(-6+14r-5r^2)\text{Li}_2(1-r) + 4(11-13r+3r^2)f(r,1) \bigg\}$$

(5.5)

Finally taking the limit $r \to 0$ one obtains the result of [4]

$$\hat{\Delta}_{lead}^{(2)}\bigg|_{h=r=0} = -N_c x_t^2 (2\pi^2 - 19) \qquad (5.6)$$

where the leading contributions of a very heavy quark were calculated with all other masses neglected.

We now present our result for $\hat{\Delta}_{sub}^{(2)}$ in full analytic form. We decompose the result as

$$\hat{\Delta}_{sub}^{(2)} = N_c x_t^2 \left[ \hat{\Delta}_1^{(2)} + \hat{\Delta}_2^{(2)} + \hat{\Delta}_3^{(2)} + \hat{\Delta}_4^{(2)} + \hat{\Delta}_5^{(2)} + \hat{\Delta}_6^{(2)} + \hat{\Delta}_{6ct}^{(2)} \right] \qquad (5.7)$$

and give the analytic form of each intermediate result $\hat{\Delta}_i^{(2)}$ as well. Each $\hat{\Delta}_i^{(2)}$ equals the contribution coming from the graphs shown in Fig.2–Fig.7 respectively. This grouping of graphs is dictated by their topology, the particles they contain in the loops, and the WI of Eq. (2.14).

The graphs of Fig.2 give

$$\begin{aligned}\hat{\Delta}_1^{(2)} =& 4r^2 \left[ -\frac{\pi^2}{12} - \frac{h^2}{4(h-r)^2} \ln^2 h + \left( -\frac{1}{4} + \frac{h^2}{4(h-r)^2} \right) \ln^2 r \right.\\ & -\frac{(1-h)^2}{(h-r)^2} \text{Li}_2(1-h) + \left( -1 + \frac{(1-h)^2}{(h-r)^2} \right) \text{Li}_2(1-r) \\ & \left. + \frac{(2-h)(4-h)}{2(h-r)^2} f(1,h) - \left( \frac{2-r}{2r} + \frac{2-h}{2r} \frac{h(2-r)+2r}{(h-r)^2} \right) f(1,r) \right]\end{aligned} \qquad (5.8)$$

In the difference of the self energies the graphs (2.a) and (2.$\alpha$) cancell for $M_W = M_Z$ and need not be computed. In the $h = r$ limit the above formula reduces to

$$\hat{\Delta}_6^{(2)}\bigg|_{h=r} = 4r^2 \left[ -\frac{\pi^2}{12} + \frac{1}{4(4-r)} \ln r - \frac{1}{4} \ln^2 r - \text{Li}_2(1-r) + \frac{r-2}{r} \left( \frac{1}{2} + \frac{1}{r(4-r)} \right) f(1,r) \right] \qquad (5.9)$$

For the uninteresting value $r = 4$ the above result as well as the results that follow are all regular as can be seen from the explicit values of $f(4,1)$ and $f'(4,1)$ given in the Appendix.

The contribution of the graphs of Fig.3 is given by

$$\begin{aligned}\hat{\Delta}_2^{(2)} =& 2r \left[ 1 + \frac{1}{h-r} + \left( 1 - \frac{1}{h-r} \right) \frac{h \ln h - r \ln r}{h-r} \right.\\ & \left( \frac{h(4-h)}{(h-r)^2} - \frac{h^2 - 6h + 6}{h-r} \right) f(1,h) + \left( \frac{-r^2 + 2rh - 6h + 2r}{(h-r)^2} + \frac{r^2 - 6r + 6}{h-r} \right) f(1,r) \\ & \left. \left( \frac{(1-h)^2}{(h-r)^2} + \frac{(3-h)(1-h)}{h-r} \right) \text{Li}_2(1-h) + \left( 1 - \frac{(1-h)^2}{(h-r)^2} - \frac{(3-r)(1-r)}{h-r} \right) \text{Li}_2(1-r) \right]\end{aligned} \qquad (5.10)$$

When $h = r$ this reduces to

$$\hat{\Delta}_2^{(2)}\bigg|_{h=r} = 2r\left[2 + \left(2 + \frac{1}{4-r}\right)\ln r + (2r-3)\text{Li}_2(1-r) + \left(7 - 2r - \frac{2}{r} + \frac{2}{r(4-r)}\right)f(1,r)\right] \tag{5.11}$$

The graphs of Fig.4 give

$$\hat{\Delta}_3^{(2)} = r\left[-\frac{12}{4-r}\ln r + r\ln^2 r + \frac{\pi^2}{3}r + 4(2+r)\text{Li}_2(1-r) + 2\left(-2 + \frac{7}{r} - r - \frac{3}{4-r}\right)f(r,1)\right] \tag{5.12}$$

Graphs (4.a) and (4.$\alpha$) as well as (4.c) with (4.$\gamma$) are equal for $M_Z = M_W$ and cancell in the difference.

From the graphs of Fig.5 we obtain

$$\hat{\Delta}_4^{(2)} = 2r\left[8 + 6\left(-1 + \frac{1}{4-r}\right)\ln r - 2r\ln^2 r - \frac{2\pi^2}{3}r - 6\text{Li}_2(1-r) + \left(14 - 4r - \frac{1}{r} + \frac{3}{4-r}\right)f(r,1)\right] \tag{5.13}$$

The graphs of Fig.6 give

$$\hat{\Delta}_5^{(2)} = r\left[-2r + r^2(4-r)\frac{\pi^2}{3} + 2\left(\frac{4}{4-r} - 2r\right)\ln r + r^2(4-r)\ln^2 r \right.$$
$$\left. - 4r(1 - 4r + r^2)\text{Li}_2(1-r) + 2\left(\frac{2}{4-r} - \frac{2}{r} + 8r - 6r^2 + r^3\right)f(r,1)\right] \tag{5.14}$$

From the graphs of Fig.7 we obtain

$$\hat{\Delta}_6^{(2)} = r\left[-\frac{3}{\epsilon} + 6\,\ell_m + \frac{47}{2} - 30r - r(4-7r)\frac{\pi^2}{3}\right.$$
$$+ 10\left(-3r + \frac{2}{4-r}\right)\ln r - r(4-7r)\ln^2 r + \tag{5.15}$$
$$\left. 2(-7 + 14r - r^2)\text{Li}_2(1-r) + 2\left(27 + \frac{5}{4-r} - \frac{5}{r} - 34r + 8r^2\right)f(r,1)\right]$$

where $\ell_m = \gamma_E + \ln(\pi m^2)$.

This last set of graphs requires counterterms which are solely due to the one-loop mass renormalization of the top quark. We perform the renormalization *on-shell*. The fermion two-point function $\Gamma_f$ is written as

$$\Gamma_f(p) = i(\not{p} - m_{0f}) - i\Sigma_f(p) \tag{5.16}$$

where $\Sigma_f$ is the fermion self-energy function. The superscript $G$ will denote the gauge boson's contribution to $\Sigma_f$ shown in Fig.9. It is given by

$$\Sigma_t^G(p) = \not{p} P_L \Sigma_{t,L}^G(p) \tag{5.17}$$

with

$$\Sigma_{t,L}^G(p) = \frac{g^2}{16\pi^2} \frac{2-D}{4} \left[ B_1(p^2, m_t^2, M_Z^2) + 2B_1(p^2, 0, M_W^2) \right] \tag{5.18}$$

$$B_1(p^2, m^2, M^2) = \frac{\delta}{2} - F_1(p^2, m^2, M^2) \;, \tag{5.19}$$

where

$$\delta = \frac{1}{\epsilon} - \gamma_E - \ln \pi \;, \text{and} \quad P_{R,L} = \frac{1 \pm \gamma_5}{2} \tag{5.20}$$

and

$$F_n(p^2, m^2, M^2) = \int_0^1 dx\, x^n \ln((1-x)m^2 + xM^2 - x(1-x)p^2) \;. \tag{5.21}$$

The mass counterterm is defined by

$$m_{0t} = m_t - \delta m_t \tag{5.22}$$

and is determined by the on-shell renormalization conditions

$$\Gamma_t(p) u_t(p)|_{\not{p}=m_t} = 0, \qquad \bar{u}_t(p) \Gamma_t(p)|_{\not{p}=m_t} = 0 \tag{5.23}$$

$$\begin{aligned}
\frac{\delta m_t}{m_t} &= \frac{1}{2} \Sigma_t^G(p^2 = m_t^2) \\
&= \frac{g^2}{32\pi^2} \left[ -\frac{3}{4}\left(\frac{1}{\epsilon} - [\gamma_E + \ln(\pi m_t^2)]\right) - 1 + \frac{3}{4}r \right. \\
&\quad \left. - \frac{5r^2 - 12r}{8} \ln r + \frac{(1-r)^2}{2} \ln|1-r| - \frac{r-2}{8}\sqrt{r} g(r) \right] \;.
\end{aligned} \tag{5.24}$$

This renormalization of the top mass gives rise to two kinds of counterterms for $\hat{\Delta}_{sub}^{(2)}$. The mass-insertion counterterms shown in Fig.7(ct1,ct3) and the vertex counterterms shown in Fig.7(ct2,ct4). The mass insertion counterterms diverge as $1/\epsilon$, which cancells in the difference, and they give a finite contribution to $\hat{\Delta}_{sub}^{(2)}$. In terms of scalar integrals this contribution equals

$$\hat{\Delta}_{6ct(mi)} = -i\frac{g^2}{(2\pi)^4}\frac{m^2}{M^2}\left[\frac{\delta m}{m}\right]\left[2\frac{(2-D)}{D}\left((mm) - \frac{(m)}{m^2}\right) + m^2(mmm)\right] , \qquad (5.25)$$

where henceforth $m \equiv m_t$. The vertex counterterms turn out to be just the one loop fermion graphs multiplied by the factor $\frac{\delta m}{m}$. In terms of scalar integrals they are given by

$$\hat{\Delta}_{6ct(v)} = -i\frac{g^2}{(2\pi)^4}\frac{m^2}{M^2}\left[\frac{\delta m}{m}\right]\left[2\frac{(2-D)}{D}\frac{(m)}{m^2} + (mm)\right] . \qquad (5.26)$$

To see how this comes about, we note that upon renormalization the $\phi\bar{\psi}\psi$ (and similarly the $\chi\bar{\psi}\psi$ ) vertex will get modified to

$$g_0\frac{m_0}{M_0}\phi_0\bar{\psi}\psi \quad \longrightarrow \quad Z_g g\frac{(m+\delta m)}{(M+\delta M)}Z_\phi^{1/2}\phi\bar{\psi}\psi ,$$

where we have ommited wave function renormalization of the fermion fields, since it will cancell agianst the corresponding wave function renormalization of the fermion propagators inside the loop. Then, at one loop the modification reads

$$1 + \delta Z_g + \frac{\delta m}{m} - \frac{\delta M}{M} + \frac{1}{2}\delta Z_\phi = 1 + \frac{\delta m}{m} ,$$

where in the last step we have used Eq. (2.20). Finally the result for the counterterms is

$$\Delta_{6ct}^{(2)} = r\left[\frac{3}{\epsilon} - 6\ell_m + \frac{5}{2} - 3r - r\left(6 - \frac{5}{2}r\right)\ln r - 2(1-r)^2\ln(1-r) \right. \\ \left. + \frac{1}{2}(r-2)\sqrt{r}g(r)\right] \qquad (5.27)$$

We note that after the inclusion of the counterterms, the terms proportional to $1/\epsilon$ present in $\Delta_6^{(2)}$ cancell, and the final result emerges finite as expected. In the limit where the mass splitting in the doublet is zero, which in our case means $m_t = m_b = 0$, all of the above expressions vanish, as can be seen from the asymptotic expressions of the functions $g(x)$, $f(x,0)$, and $f(x,1)$, if we take $x \to \infty$ (for $x = r$ or $h$).

Adding together Eq. (5.8) to Eq. (5.27) we report as our final result

$$\begin{aligned}\Delta^{(2)}_{sub} =\ & N_c x_t^2\ S(h,r) \\
=\ & N_c x_t^2\ 2r\Bigg\{22 - \frac{35}{2}r + \frac{1}{h-r} - r(8+3r+r^2)\frac{\pi^2}{6} + \frac{1}{4}(r-2)\sqrt{r}g(r) \\
& - \left[\frac{r}{h-r}\left(1-\frac{1}{h-r}\right) + 6 + 20r + \frac{5}{4}r^2 - \frac{14}{4-r}\right]\ln r + \frac{h}{h-r}\left[1-\frac{1}{h-r}\right]\ln h \\
& - (1-r)^2\ln|1-r| + \left[\frac{rh^2}{2(h-r)^2} - 4r + \frac{11}{2}r^2 - \frac{1}{2}r^3\right]\ln^2 r - \frac{rh^2}{2(h-r)^2}\ln^2 h \\
& - \left[(1-2r)\frac{(1-h)^2}{(h-r)^2} + \frac{(3-r)(1-r)}{h-r} + 8 - 12r - 7r^2 + 2r^3\right]\text{Li}_2(1-r) \\
& + \left[(1-2r)\frac{(1-h)^2}{(h-r)^2} + \frac{(3-h)(1-h)}{h-r}\right]\text{Li}_2(1-h) \\
& - \left[\frac{r(2-h)(4-h) + r(4-r)}{(h-r)^2} - \frac{(r-2)^2 + 2(h-4)}{h-r} - 37 + 30r + \frac{1}{r} - \frac{7}{4-r}\right]f(r,1) \\
& + \left[\frac{r(2-h)(4-h) + h(4-h)}{(h-r)^2} - \frac{(h-2)^2 + 2(1-h)}{h-r}\right]f(h,1)\Bigg\}\ .\end{aligned}$$

(5.28)

In Figure 10 we show together the functions $R(h,r)$ and $S(h,r)$, which describe respectively the leading and subleading contributions, for $M/m_t = 0.5$ or $r = 0.25$ and a wide range of values for the mass of the Higgs boson, $M_H$, where $h = M_H^2/m_t^2$. The function $R(h,r)$ of the leading contributions grows asymptoticaly in the way described in [6]. On the other hand, the function $S(h,r)$ of the subleading contributions becomes independent $M_H$ for a very heavy Higgs boson (as pointed out in [7]). For the phenomenologically interesting range of values for $M_H$, the two functions have opposite sign. In Table 1 we give numerical results for $\hat{\Delta}^{(2)}$ for different values of the ratios $M/m_t$ and $M_H/m_t$ in units of $N_c x_t^2$. The first entry of each column corresponds to the leading contribution $\hat{\Delta}^{(2)}_{lead}$, while the second gives the total correction $\hat{\Delta}^{(2)}_{lead} + \hat{\Delta}^{(2)}_{sub}$. We notice that for a light Higgs boson, where $\hat{\Delta}^{(2)}$ is small, the two contributions $\hat{\Delta}^{(2)}_{lead}$ and $\hat{\Delta}^{(2)}_{sub}$ are comparable in magnitude, as can be seen from the entries in the first three lines of the first four columns. We also observe that for a light top ($M/m_t = 0.6$, fifth column) the two corrections come with the same negative sign for $M_H \leq m_t$. On the other hand, for the largest part of the parameter space of $m_t$ and $M_H$, the subleading contributions $\hat{\Delta}^{(2)}_{sub}$ are approximately 22-27% of the leading $\hat{\Delta}^{(2)}_{lead}$ contributions, which is what one would naively expect. Finally, in Figure 11 we give the two loop correction $\hat{\Delta}^{(2)}$ as a function of $M_H$ for $M = 91.19$ GeV and $m_t = 175$ GeV.

## 6. Conclussions

In this paper we showed that in the framework of the PT one can define a universal part of the $\rho$ parameter, which satisfies all necessary field theoretical requirements, for *both* leading and sub-leading two-loop corrections. Most noticeably, the PT universal part is by construction independent of the gauge-fixing parameter, and, at the same time, process independent; furthermore, by virtue of the PT Ward identities it is also UV finite.

We have calculated the two loop contributions at the subleading order $\mathcal{O}(G_\mu^2 m_t^2 M_Z^2)$, in the limit $M_W = M_Z$, and $s = 0$. Their relative size was found to be around 25% with respect to the leading ones. From the technical point of view, the computation involved self-energy garphs only, which exist in closed analytic form. The computational part was significantly fascilitated by the PT Ward identities, relating the self-energies of the gauge bosons ($WW$ and $ZZ$) to the corresponding self-energies of the would-be Goldstone bosons ($\phi\phi$ and $\chi\chi$). These Ward identities, which are valid for *both* leading and sub-leading contributions, reduce the task into calculating Goldstone boson self-energies only.

Having laid out the framework of how such calculations should proceed, it is straightforward to compute the two loop corrections to the universal part of $\rho$ defined via the PT, for the full $SU(2)_L \times U(1)_Y$ standard model. Results in this direction will be presented elsewhere.

## 7. Acknowledgements

The authors thank P. Gambino and A. Sirlin for useful discussions. K.S. would like to thank A. Sirlin for the hospitality extended to him at the New York University, where part of this work was done. This work was supported in part by the National Science Foundation Grant No.PHY-9313781.

## 8. Figure captions

**Figure 1** : The $\xi_Q$ dependent part of $\Delta$ in the background field gauges at one loop (for $M_H = 300$ GeV and in units of $g^2/(16\pi^2)$).

**Figure 2** : Subleading two loop graphs containing self energy insertions of a Higgs boson or a gauge boson.

**Figure 3** : The subleading fermionic triangle graphs that contain a Higgs boson.

**Figure 4** : The rest of the subleading graphs containing bosonic self energy insertions.

**Figure 5** : The rest of the subleading triangle graphs.

**Figure 6** : Subleading fermionic bubble graphs with a vertex correction.

**Figure 7** : Subleading fermionic bubble graphs with self energy insertions and their relevant counterterms.

**Figure 8** : Subleading graphs that contain a mixed self energy insertion. The graphs that correspond to the scalar self energy $\phi\phi$ vanish for $\sin\theta_W = 0$.

**Figure 9** : The gauge boson contributions to the one loop top self energy.

**Figure 10** : The functions $R(h, r)$ and $S(h, r)$ describing respectively the leading and the subleading contributions to $\hat{\Delta}$, for $r = 0.25$.

**Figure 11** : The universal two loop $\hat{\Delta}^{(2)}$ correction to the $\rho$ parameter as a function of the Higgs boson mass, for different approximations, in units of $N_c x_t^2$ . Dotted line : the leading result in the approximation $M_W = M_Z = 0$, [5], [6]. Dashed line : The leading contribution in the approximation $M_W = M_Z = M$ [7]. Solid line : the total correction, leading and subleading, for $M_W = M_Z = M$. ($m_t = 175$ GeV, $M = 91.19$ GeV).

**Table 1** : Numerical values for $\hat{\Delta}^{(2)}$ in units of $N_c x_t^2$, for different values of $\sqrt{h} = M_H/m_t$ and $\sqrt{r} = M/m_t$. The first entry of each column corresponds to the leading contribution $\hat{\Delta}^{(2)}_{lead}$ in the approximation $M_W = M_Z = M$, while the second gives the total result $\hat{\Delta}^{(2)} = \hat{\Delta}^{(2)}_{lead} + \hat{\Delta}^{(2)}_{sub}$ in the same approximation.

# Appendix I.
# Scalar two loop integrals

The scalar two loop integrals at zero external momentum are defined as in [39] and [40]:

$$(\underbrace{m_1...m_1}_{\nu_1} | \underbrace{m_2...m_2}_{\nu_2} | \underbrace{m_3...m_3}_{\nu_3}) = \int\int \frac{d^D k\, d^D \ell}{[k^2 - m_1^2]^{\nu_1} [\ell^2 - m_2^2]^{\nu_2} [(k+\ell)^2 - m_3^2]^{\nu_3}} \quad (I.1)$$

Obviously

$$(\underbrace{m_1...m_1}_{\nu_1} | \underbrace{m_2...m_2}_{\nu_2} | \underbrace{m_3...m_3}_{\nu_3}) = (\underbrace{m_2...m_2}_{\nu_2} | \underbrace{m_1...m_1}_{\nu_1} | \underbrace{m_3...m_3}_{\nu_3}) = (\underbrace{m_3...m_3}_{\nu_3} | \underbrace{m_2...m_2}_{\nu_2} | \underbrace{m_1...m_1}_{\nu_1})$$

Note that, within each of the three groups of masses in Eq. (I.1), all masses are equal. Any other integral, for which different masses appear within the same group, can be brought in the form of Eq. (I.1) by use of partial fractions

$$\frac{1}{k^2 - m_1^2} \frac{1}{k^2 - m_2^2} = \frac{1}{m_1^2 - m_2^2}\left[\frac{1}{k^2 - m_1^2} - \frac{1}{k^2 - m_2^2}\right]. \quad (I.2)$$

All of these integrals can be calculated from the knowledge of a single master integral, either by use of recurrence relations, derived through integration by parts, or by differentiation with respect to one of the masses. For example

$$(MMM|M_1|M_2) = \frac{1}{2}\frac{\partial}{\partial M^2}(MM|M_1|M_2) \quad (I.3)$$

and

$$(MM|M_1 M_1|M_2) = \frac{\partial}{\partial M_1^2}(MM|M_1|M_2) \ . \quad (I.4)$$

For our purposes, we have found the following two recursion relations very useful

$$\begin{aligned}(M_1)(M_3 M_3) - (M_3)(M_1 M_1) + (M_2)\left((M_1 M_1) - (M_3 M_3)\right) = \\ (M_1^2 - M_2^2 - M_3^2)(M_3 M_3|M_1|M_2) + (M_1^2 + M_2^2 - M_3^2)(M_1 M_1|M_2|M_3)\end{aligned} \quad (I.5)$$

and
$$(D-3)(M_1|M_2|M_3) = M_1^2(M_1M_1|M_2|M_3) + M_2^2(M_2M_2|M_1|M_3) \\ + M_3^2(M_3M_3|M_1|M_2) ,\qquad(I.6)$$

The integrals of the type $(MM)$, are one loop integrals and are defined at the end of this Appendix. To prove Eq. (I.5) we consider

$$S(q^2) \equiv \int d^D k\, d^D \ell \frac{2k \cdot q}{[(\ell+q)^2 - m_1^2](k^2 - m_2^2)[(\ell+q+k)^2 - m_3^2]} \qquad(I.7)$$

which is zero as can be easily seen by the shift of variables $\ell \to \ell + q$. Then from

$$\left.\frac{dS(q^2)}{dq^2}\right|_{q^2=0} \equiv \frac{1}{2D}\left(\Box_q S(q^2)\right)_{q=0} = 0 \qquad(I.8)$$

one obtains Eq. (I.5). To prove the identity of Eq. (I.6), which has also been given in [39], we use

$$L(q^2) \equiv \int d^D k\, d^D \ell \frac{\partial}{\partial k^\mu}\left[\frac{k_\mu}{(k^2 - M_1^2)(\ell^2 - M_2^2)[(k+\ell)^2 - M_3^2]}\right] = 0 \qquad(I.9)$$

and Eq. (I.5).

In [39], the master integral $(MM|M_1|M_2)$ has been calculated. In [40] the same integrals have been discussed in a different mathematical framework and the master integral $(M|M_1|M_2)$ is given. In what follows we use the notation of [39]. With $D = 4 - 2\epsilon$ the result for the master integral is :

$$(MM|M_1|M_2) = \pi^4 \left[-\frac{1}{2\epsilon^2} - \frac{1}{2\epsilon}(1 - 2\ell_M) + \ell_M - \ell_M^2 - \frac{1}{2} - \frac{\pi^2}{12} - f(a,b)\right] \qquad(I.10)$$

where

$$\ell_M = \gamma_E + \ln \pi M^2, \quad a = \frac{M_1^2}{M^2}, \quad b = \frac{M_2^2}{M^2} .$$

The real function $f(a,b)$ is symmetric in its arguments and defined as

$$f(a,b) = \int_0^1 dx \left(\text{Li}_2(1 - \mu^2) - \frac{\mu^2}{1 - \mu^2}\ln \mu^2\right), \quad \mu^2 = \frac{ax + b(1-x)}{x(1-x)} ,$$

with $\text{Li}_2$ the dilogarithm function

$$\text{Li}_2(x) = -\int_0^x dy \frac{\ln(1-y)}{y} \qquad(I.11)$$

The explicit form of $f(a,b)$ is

$$f(a,b) = -\frac{1}{2}\ln a \ln b + \frac{1-a-b}{\sqrt{\phantom{x}}}\left[\frac{\pi^2}{6} + \text{Li}_2\left(-\frac{x_2}{y_1}\right) + \text{Li}_2\left(-\frac{y_2}{x_1}\right)\right.$$
$$\left. + \frac{1}{4}\ln^2\frac{x_2}{y_1} + \frac{1}{4}\ln^2\frac{y_2}{x_1} + \frac{1}{4}\ln^2\frac{x_1}{y_1} - \frac{1}{4}\ln^2\frac{x_2}{y_2}\right] \quad (I.12)$$

with

$$\sqrt{\phantom{x}} = \sqrt{1 - 2(a+b) + (a-b)^2}, \quad x_{1,2} = \frac{1}{2}\left[1 + b - a \pm \sqrt{\phantom{x}}\right], \quad y_{1,2} = \frac{1}{2}\left[1 + a - b \pm \sqrt{\phantom{x}}\right].$$

It turns out that all of our results can be expressed in terms of $f(a,0)$ and $f(a,1)$. These functions are explicitly given by:

$$f(a,0) = \text{Li}_2(1-a) = \frac{\pi^2}{6} - \text{Li}_2(a) - \ln a \ln(1-a), \quad (I.13)$$

and

$$f(a,1) = -\frac{1}{\sqrt{4-y}}\left[\frac{\pi^2}{6} + 2\text{Li}_2(\zeta) + \frac{1}{2}\ln^2\zeta\right] \quad 4 < a$$
$$= -4\ln 2 \quad a = 4 \quad (I.14)$$
$$= -\frac{2}{\sqrt{y-4}}\text{Cl}_2(\varphi) \quad a > 4$$

where $\text{Cl}_2(x)$ is the Clausen function

$$\text{Cl}_2(x) = \text{Im}[\text{Li}_2(e^{ix})], \quad (I.15)$$

and the variables $y$, $\zeta$, and $\phi$ are defined as in [6]

$$y = \frac{4}{x}, \quad \zeta = \frac{\sqrt{1-y}-1}{\sqrt{1-y}+1}, \quad \text{and } \phi = 2\arcsin\left(\frac{\sqrt{x}}{2}\right).$$

The derivatives of $f(a,1)$ and $f(a,0)$ are given by

$$\frac{df(a,1)}{da} = \frac{1}{4-a}\left[\ln a + \frac{2}{a}f(a,1)\right], \quad a \neq 4,$$
$$= -\frac{1}{6}(1 + 2\ln 2), \quad a = 4$$

$$\frac{df(a,0)}{da} = \frac{\ln a}{1-a}$$

We will also need the following derivative

$$\left.\frac{\partial f(a,b)}{\partial b}\right|_{b=1} = -\frac{2}{4-a}\left[\ln a + \left(1-\frac{2}{a}\right)f(a,1)\right] , \quad a \neq 4$$

$$= \frac{1}{3} - \frac{4}{3}\ln 2 , \qquad a = 4 .$$

which we obtain directly from Eq. (I.12).

The following relationships are useful in expressing the results in terms of only these two functions

$$f(\frac{1}{a},0) = -f(a,0) - \frac{1}{2}\ln^2 a$$

$$f(\frac{1}{a},\frac{1}{a}) = -\left(1-\frac{2}{a}\right)f(a,1) - \frac{1}{2}\ln^2 a .$$
(I.16)

In our calculations we have encountered two loop integrals with three, four, and five propagators. All integrals with three propagators are converted to integrals with four propagators using Eq. (I.6). Most of the integrals with four propagators can be directly obtained from Eq. (I.10). For the integrals $(00|m|M)$ and $(00|0|M)$ we use Eq. (I.5), and write them as

$$(1-r)(00|m|M) = \frac{1}{m^2}(m)(MM) - (1+r)(MM|m|0)$$

and

$$(00|0|M) = -(MM|0|0)$$

with $r = M^2/m^2$ as usual.

The integrals with five propagators that we encountered are the following :

$$(MMM|m|m) = \frac{1}{2}\frac{\partial}{\partial M^2}(MM|m|m)$$
$$= \frac{\pi^4}{2M^2}\left[\frac{1}{\epsilon} + 1 - 2\ell_M + \frac{2\ln r}{4-r} + \frac{4}{r(4-r)}f(r,1)\right]$$
$$(MMM|m|0) = \frac{1}{2}\frac{\partial}{\partial M^2}(MM|m|0)$$
$$= \frac{\pi^4}{2M^2}\left[\frac{1}{\epsilon} + 1 - 2\ell_M + \frac{\ln r}{1-r}\right]$$
$$(MMM|0|0) = \frac{1}{2}\frac{\partial}{\partial M^2}(MM|0|0)$$
$$= \frac{\pi^4}{2M^2}\left[\frac{1}{\epsilon} + 1 - 2\ell_M\right]$$
(I.17)

$$(mmm|m|M) = \frac{1}{2}\frac{\partial}{\partial m^2}(mm|\mu|M)\bigg|_{\mu=m}$$

$$= \frac{\pi^4}{2m^2}\left[\frac{1}{\epsilon} + 1 - 2\ell_m + \frac{\partial f(a,r)}{\partial a}\bigg|_{a=1} + r\frac{\partial f(1,r)}{\partial r}\right]$$

$$= \frac{\pi^4}{2m^2}\left[\frac{1}{\epsilon} + 1 - 2\ell_m - \frac{2-r}{4-r}\ln r + \frac{4}{r(4-r)}f(r,1)\right]$$

$$(mm|mm|M) = \frac{\partial}{\partial \mu^2}(mm|\mu|M)\bigg|_{\mu=m} = -\frac{\pi^4}{m^2}\frac{\partial f(r,b)}{\partial b}\bigg|_{b=1} \qquad (I.18)$$

$$= \frac{\pi^4}{m^2}\frac{2}{4-r}\left[\ln r + \left(1 - \frac{2}{r}\right)f(r,1)\right]$$

$$(MM|mm|0) = \frac{\partial}{\partial m^2}(MM|m|0) = \frac{\partial}{\partial M^2}(mm|M|0) = -\frac{\pi^4}{m^2}\frac{df(r,1)}{dr}$$

$$= -\frac{\pi^4}{m^2}\frac{\ln r}{1-r}$$

Besides the two loop integrals our results also contain products of one loop integrals. These integrals can all be computed from the integral $(m)$ through differentiation.

$$(\underbrace{m...m}_{n}) \equiv \int \frac{d^n k}{[k^2 - m^2]^n} = \frac{1}{(n-1)!}\frac{d^n}{d(m^2)^n}(m)$$

The expression for $(m)$ up to $\mathcal{O}(\epsilon)$ is

$$(m) = \int \frac{d^n k}{k^2 - m^2} = i\pi m^2\left[\frac{1}{\epsilon} + 1 - \ell_m + \epsilon\left(1 + \frac{\pi^2}{12} - \ell_m + \frac{1}{2}\ell_m^2\right)\right] . \qquad (I.19)$$

# Appendix II.
# Feynman rules in the background field gauges

In this appendix we give the Feynman rules that are relevant for the calculation of the subleading two loop contributions to $\hat{\Delta}$. The rules are given for the full standard model, and the calculations in the text were carried out in the approximation : $M_W = M_Z$, $s = 0$, $m_b = 0$.

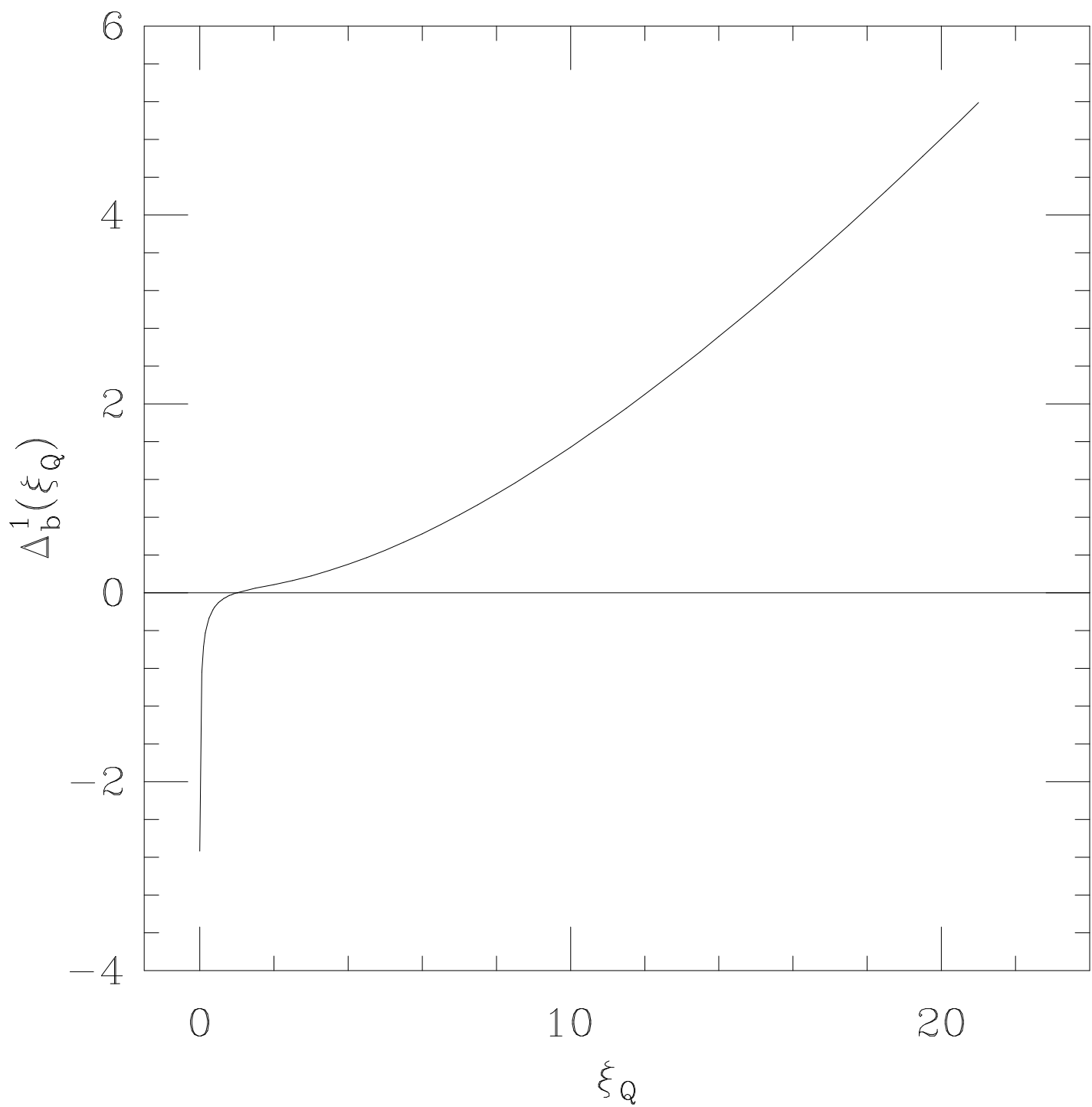

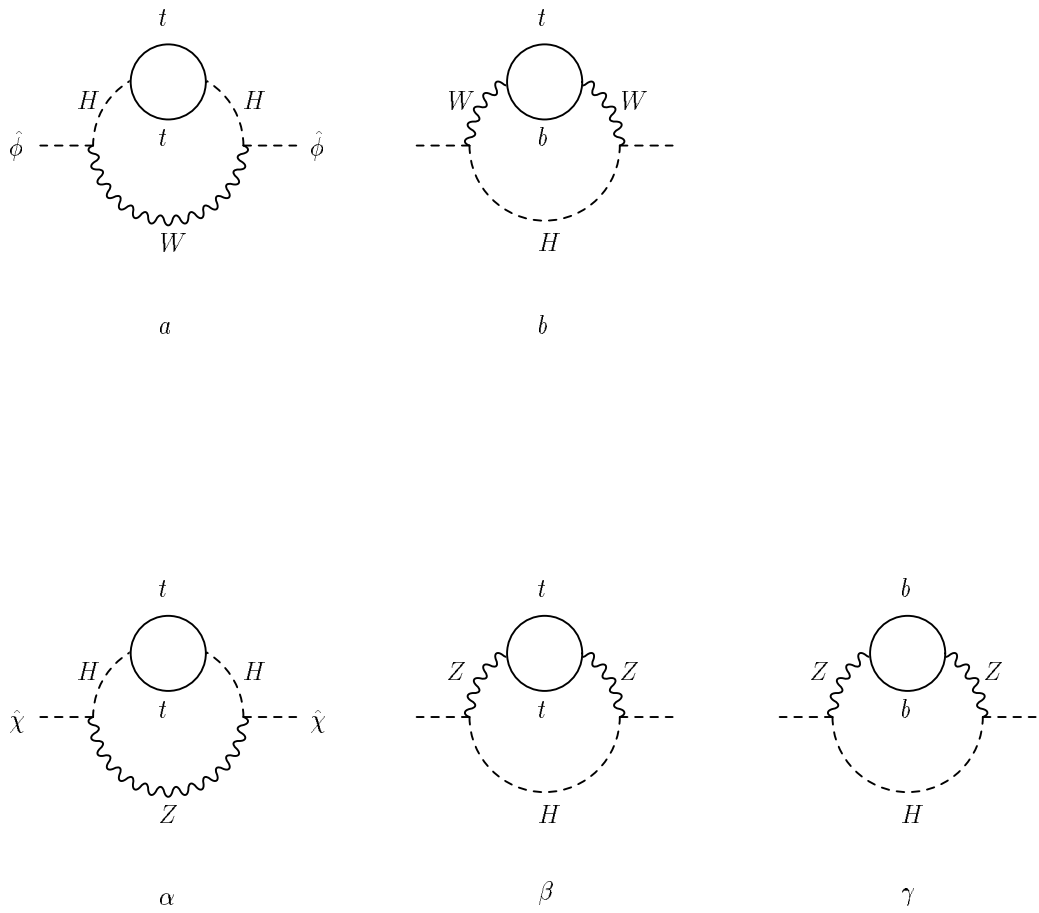

Figure 2

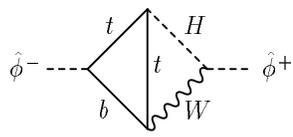 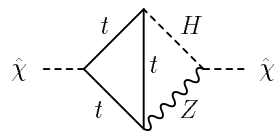 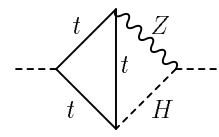

Figure 3

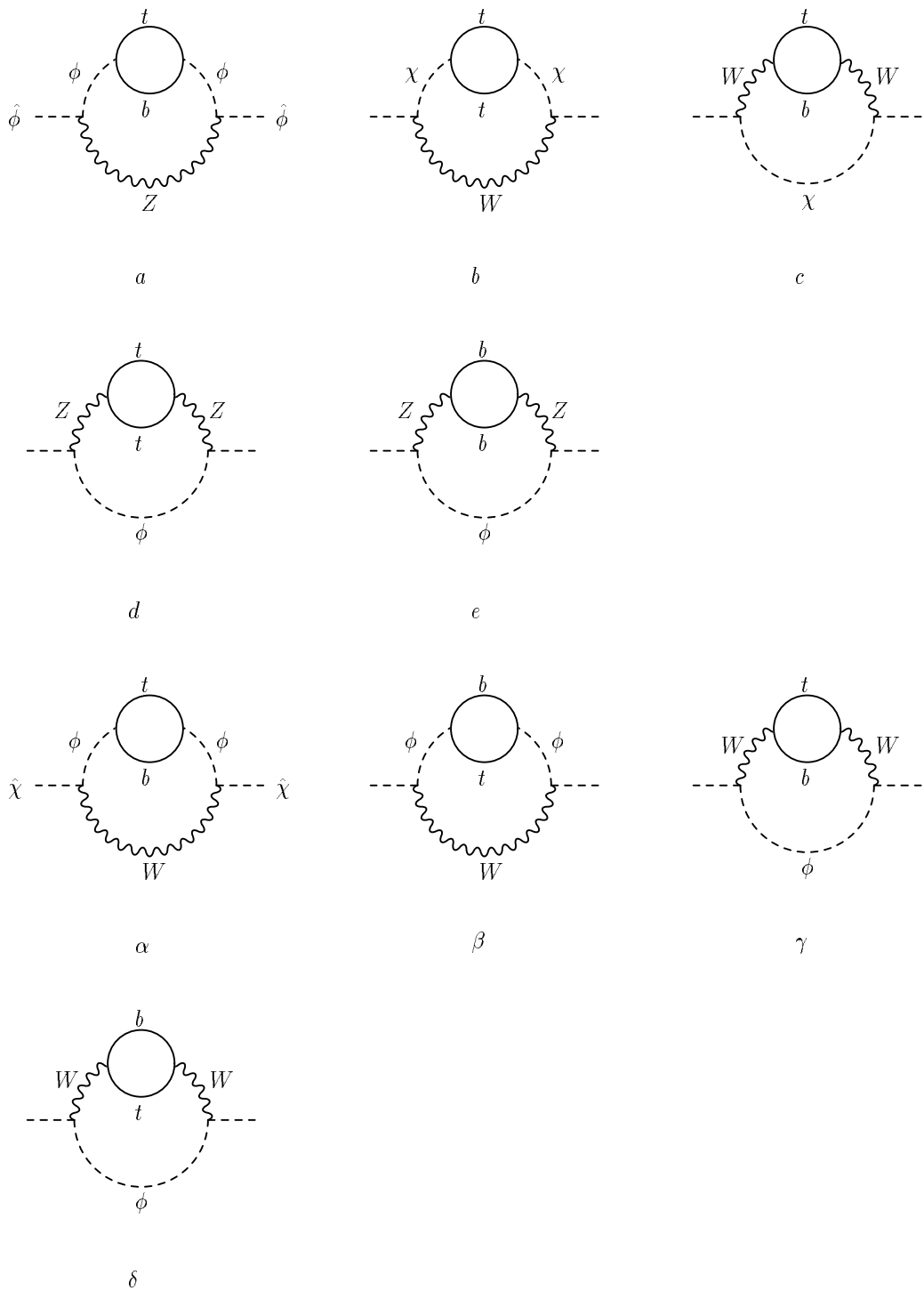

Figure 4

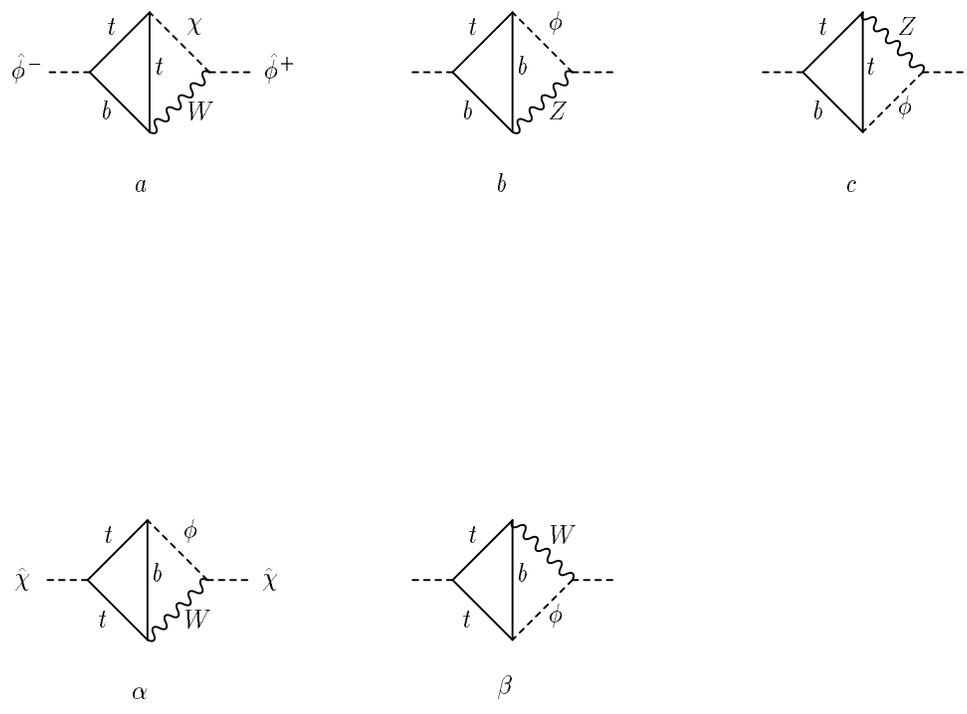

Figure 5

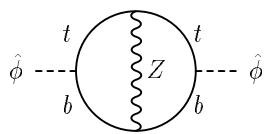 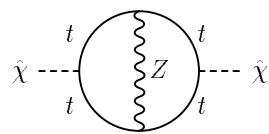

a  α

Figure 6

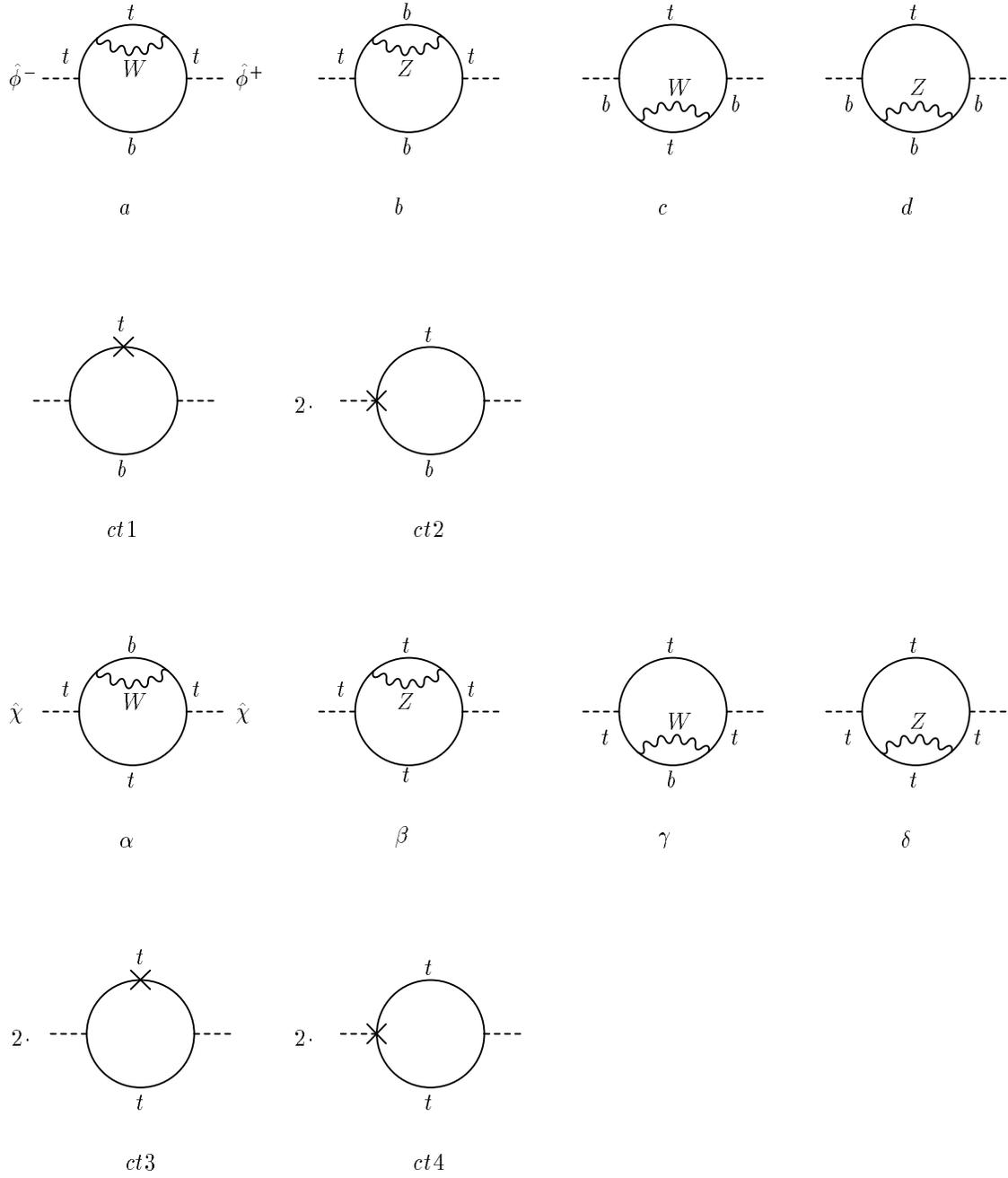

Figure 7

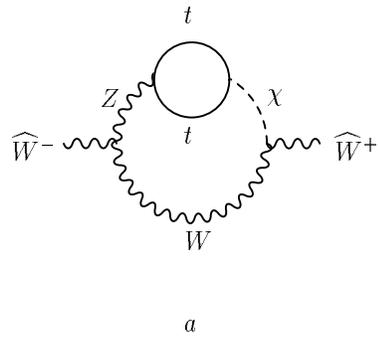
a

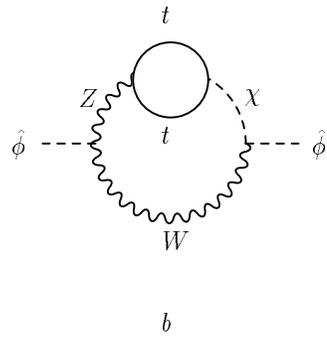
b

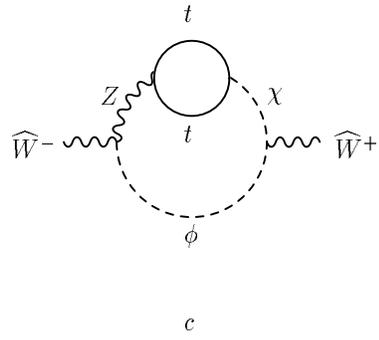
c

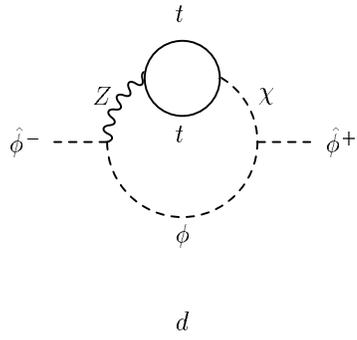
d

Figure 8

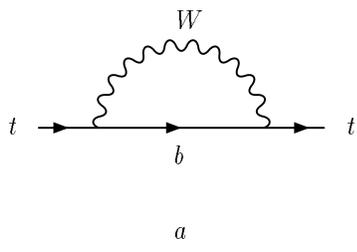 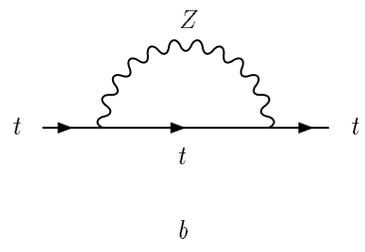

Figure 9

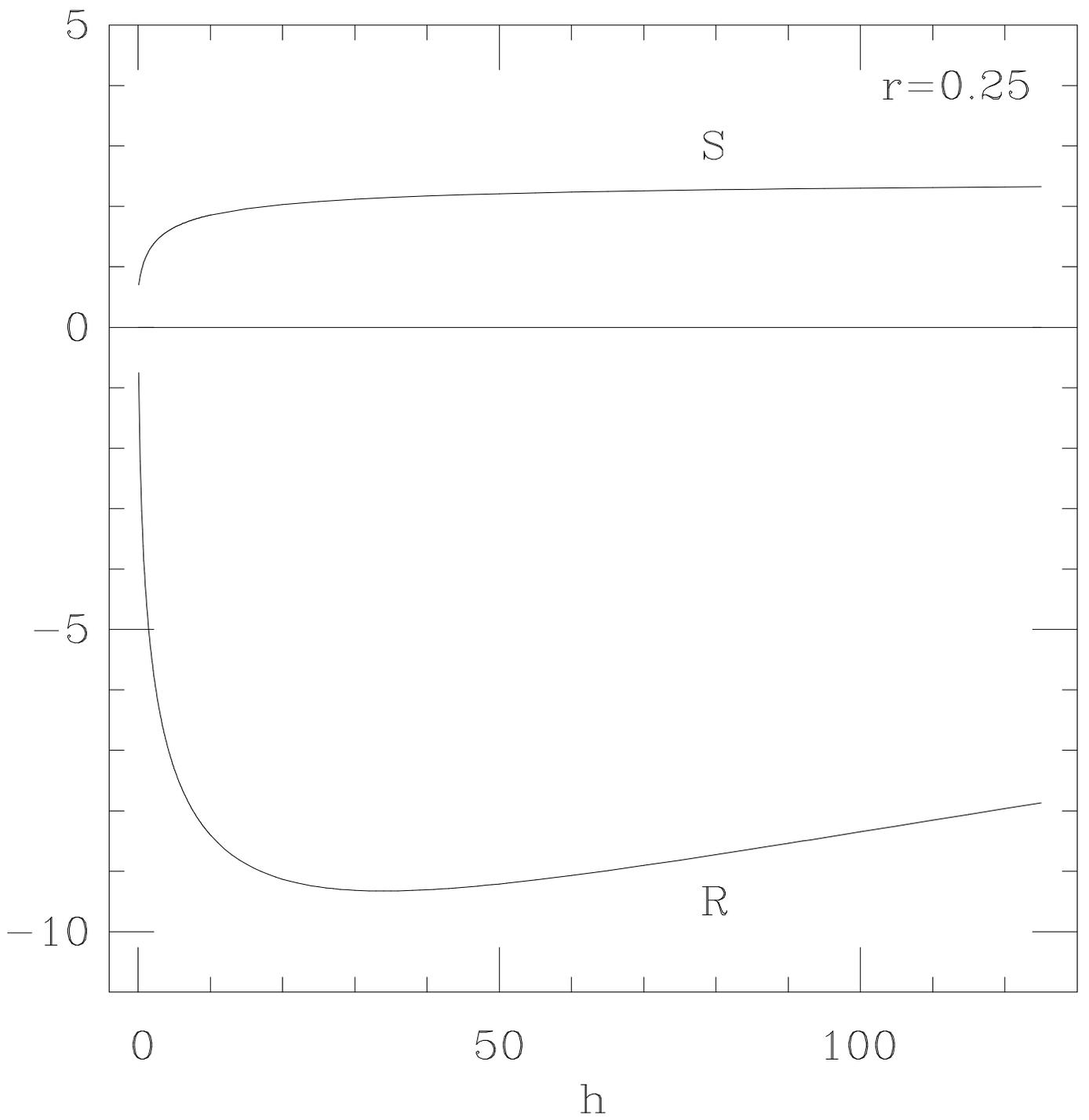

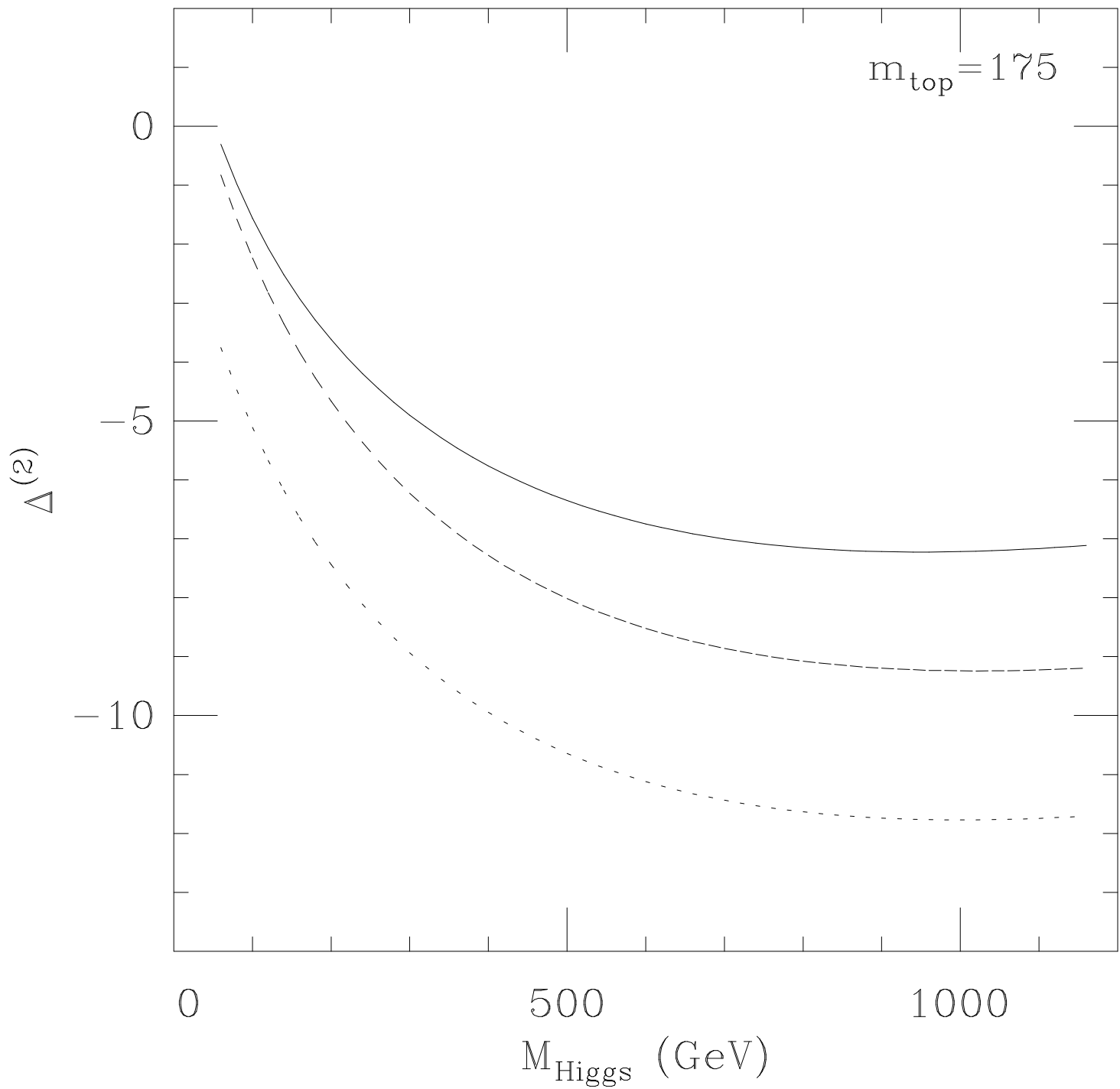

| $M_H/m_t$ | $M/m_t$ | | | | | | | | | |
|---|---|---|---|---|---|---|---|---|---|---|
| | 0.40 | | 0.45 | | 0.50 | | 0.55 | | 0.60 | |
| | leading | total | leading | total | leading | total | leading | total | leading | total |
| 0.4 | −1.926 | −0.691 | −1.617 | −0.540 | −1.328 | −0.576 | −1.062 | −0.833 | −0.822 | −1.345 |
| 0.6 | −3.083 | −1.737 | −2.782 | −1.582 | −2.500 | −1.615 | −2.243 | −1.871 | −2.012 | −2.383 |
| 0.8 | −4.032 | −2.581 | −3.737 | −2.417 | −3.462 | −2.443 | −3.212 | −2.693 | −2.989 | −3.201 |
| 1.0 | −4.828 | −3.284 | −4.538 | −3.110 | −4.270 | −3.127 | −4.026 | −3.369 | −3.809 | −3.869 |
| 1.2 | −5.506 | −3.880 | −5.221 | −3.697 | −4.957 | −3.704 | −4.718 | −3.937 | −4.508 | −4.429 |
| 1.4 | −6.089 | −4.393 | −5.808 | −4.199 | −5.548 | −4.198 | −5.315 | −4.421 | −5.109 | −4.905 |
| 1.6 | −6.595 | −4.837 | −6.317 | −4.635 | −6.061 | −4.624 | −5.832 | −4.839 | −5.631 | −5.315 |
| 1.8 | −7.036 | −5.223 | −6.761 | −5.013 | −6.508 | −4.994 | −6.282 | −5.202 | −6.086 | −5.669 |
| 2.0 | −7.421 | −5.562 | −7.149 | −5.344 | −6.900 | −5.318 | −6.677 | −5.518 | −6.484 | −5.978 |
| 2.2 | −7.760 | −5.858 | −7.490 | −5.634 | −7.243 | −5.601 | −7.024 | −5.794 | −6.834 | −6.247 |
| 2.4 | −8.057 | −6.118 | −7.789 | −5.888 | −7.545 | −5.849 | −7.328 | −6.036 | −7.141 | −6.483 |
| 2.6 | −8.319 | −6.346 | −8.053 | −6.111 | −7.811 | −6.066 | −7.597 | −6.247 | −7.412 | −6.688 |
| 2.8 | −8.549 | −6.546 | −8.285 | −6.306 | −8.045 | −6.256 | −7.833 | −6.432 | −7.651 | −6.868 |
| 3.0 | −8.751 | −6.721 | −8.488 | −6.476 | −8.251 | −6.422 | −8.040 | −6.593 | −7.861 | −7.024 |
| 3.2 | −8.928 | −6.874 | −8.667 | −6.625 | −8.431 | −6.566 | −8.223 | −6.733 | −8.045 | −7.159 |
| 3.4 | −9.082 | −7.006 | −8.823 | −6.754 | −8.589 | −6.691 | −8.382 | −6.853 | −8.207 | −7.276 |
| 3.6 | −9.217 | −7.121 | −8.959 | −6.865 | −8.726 | −6.798 | −8.521 | −6.957 | −8.348 | −7.376 |
| 3.8 | −9.334 | −7.219 | −9.077 | −6.960 | −8.845 | −6.890 | −8.642 | −7.045 | −8.470 | −7.460 |
| 4.0 | −9.434 | −7.302 | −9.178 | −7.040 | −8.948 | −6.967 | −8.746 | −7.119 | −8.576 | −7.531 |
| 4.2 | −9.519 | −7.372 | −9.265 | −7.107 | −9.035 | −7.031 | −8.835 | −7.180 | −8.666 | −7.589 |
| 4.4 | −9.591 | −7.429 | −9.337 | −7.162 | −9.109 | −7.083 | −8.910 | −7.230 | −8.742 | −7.635 |
| 4.6 | −9.650 | −7.475 | −9.397 | −7.205 | −9.170 | −7.124 | −8.972 | −7.268 | −8.806 | −7.671 |
| 4.8 | −9.698 | −7.510 | −9.446 | −7.238 | −9.220 | −7.156 | −9.023 | −7.297 | −8.858 | −7.698 |
| 5.0 | −9.735 | −7.536 | −9.485 | −7.262 | −9.259 | −7.178 | −9.063 | −7.317 | −8.899 | −7.715 |
| 5.2 | −9.763 | −7.553 | −9.513 | −7.278 | −9.289 | −7.191 | −9.094 | −7.328 | −8.931 | −7.724 |
| 5.4 | −9.782 | −7.562 | −9.533 | −7.285 | −9.310 | −7.196 | −9.116 | −7.332 | −8.953 | −7.725 |
| 5.6 | −9.793 | −7.563 | −9.544 | −7.285 | −9.322 | −7.195 | −9.129 | −7.328 | −8.968 | −7.720 |
| 5.8 | −9.796 | −7.558 | −9.548 | −7.278 | −9.327 | −7.186 | −9.134 | −7.318 | −8.974 | −7.708 |
| 6.0 | −9.793 | −7.546 | −9.545 | −7.264 | −9.324 | −7.171 | −9.133 | −7.301 | −8.973 | −7.690 |
| 6.2 | −9.782 | −7.528 | −9.536 | −7.245 | −9.315 | −7.150 | −9.125 | −7.279 | −8.966 | −7.666 |
| 6.4 | −9.766 | −7.505 | −9.520 | −7.220 | −9.300 | −7.124 | −9.110 | −7.252 | −8.952 | −7.637 |
| 6.6 | −9.744 | −7.476 | −9.499 | −7.191 | −9.280 | −7.093 | −9.090 | −7.219 | −8.933 | −7.603 |

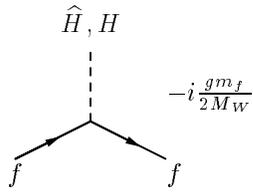

$\widehat{H}, H$

$-i\frac{gm_f}{2M_W}$

$f \qquad f$

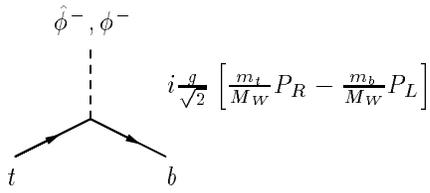

$\hat{\phi}^-, \phi^-$

$i\frac{g}{\sqrt{2}}\left[\frac{m_t}{M_W}P_R - \frac{m_b}{M_W}P_L\right]$

$t \qquad b$

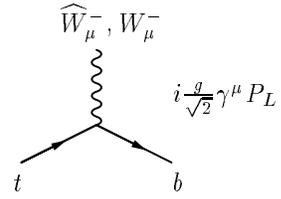

$\widehat{W}_\mu^-, W_\mu^-$

$i\frac{g}{\sqrt{2}}\gamma^\mu P_L$

$t \qquad b$

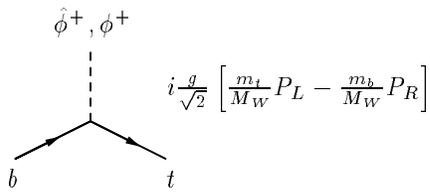

$\hat{\phi}^+, \phi^+$

$i\frac{g}{\sqrt{2}}\left[\frac{m_t}{M_W}P_L - \frac{m_b}{M_W}P_R\right]$

$b \qquad t$

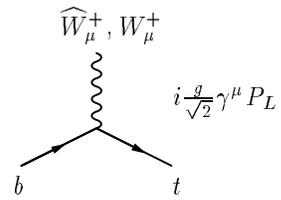

$\widehat{W}_\mu^+, W_\mu^+$

$i\frac{g}{\sqrt{2}}\gamma^\mu P_L$

$b \qquad t$

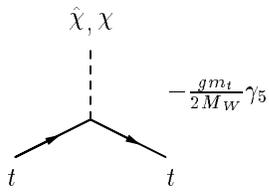

$\hat{\chi}, \chi$

$-\frac{gm_t}{2M_W}\gamma_5$

$t \qquad t$

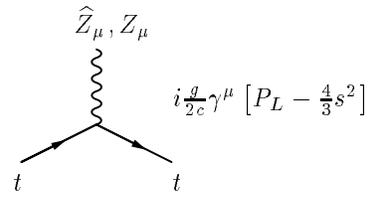

$\widehat{Z}_\mu, Z_\mu$

$i\frac{g}{2c}\gamma^\mu\left[P_L - \frac{4}{3}s^2\right]$

$t \qquad t$

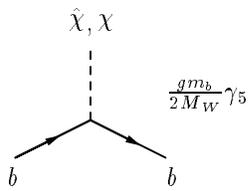

$\hat{\chi}, \chi$

$\frac{gm_b}{2M_W}\gamma_5$

$b \qquad b$

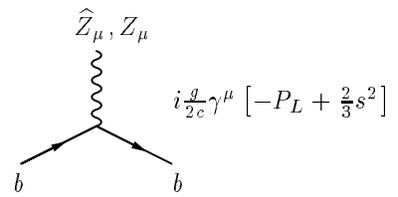

$\widehat{Z}_\mu, Z_\mu$

$i\frac{g}{2c}\gamma^\mu\left[-P_L + \frac{2}{3}s^2\right]$

$b \qquad b$

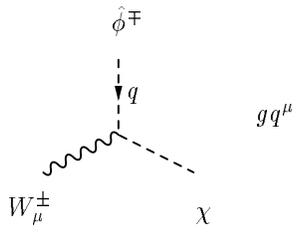 $gq^\mu$

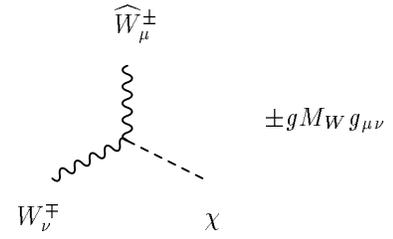 $\pm g M_W g_{\mu\nu}$

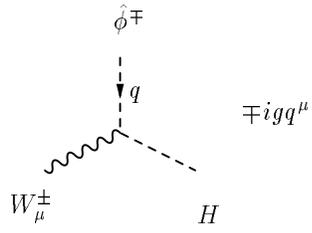 $\mp ig q^\mu$

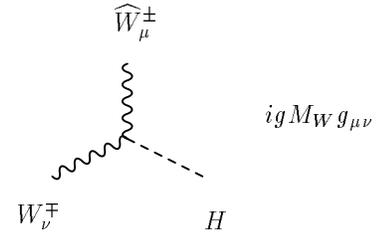 $ig M_W g_{\mu\nu}$

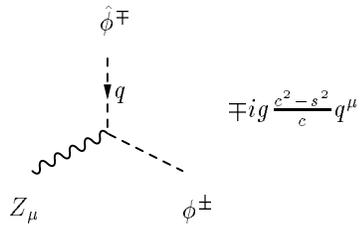 $\mp ig \frac{c^2-s^2}{c} q^\mu$

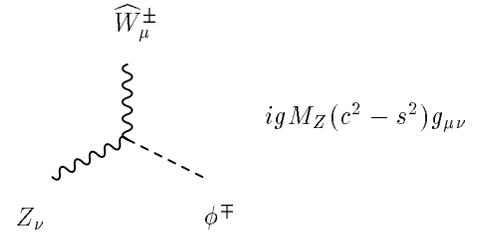 $ig M_Z (c^2 - s^2) g_{\mu\nu}$

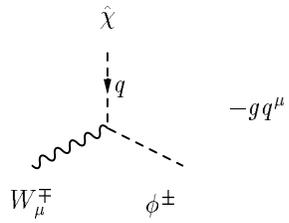 $-g q^\mu$

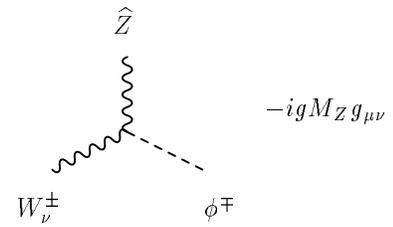 $-ig M_Z g_{\mu\nu}$

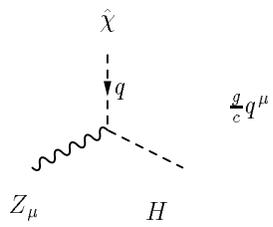 $\frac{g}{c} q^\mu$

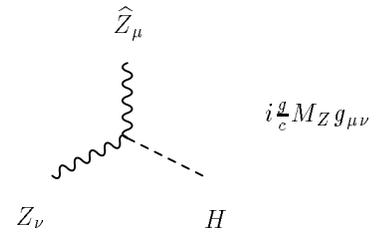 $i\frac{g}{c} M_Z g_{\mu\nu}$